\documentclass[useAMS,usenatbib,usegraphicx]{mn2e}

\newcommand{\smfont}[1]{\textnormal{{\tiny #1}}}

\title[Pulsar Galactic dynamics]{Populating the Galaxy with Pulsars -- II. Galactic dynamics}
\author[P.D. Kiel and J.R. Hurley]{Paul D. Kiel$^{1}$\thanks{E-mail:
pkiel@astro.swin.edu.au (PDK)} and Jarrod R. Hurley$^{1}$\\
$^{1}$Centre for Astrophysics and Supercomputing, Swinburne University of Technology, Hawthorn, Victoria, 3122, Australia}

\begin{document}

\date{Accepted xxx. Received xxx; in original form xxx}

\pagerange{\pageref{firstpage}--\pageref{lastpage}} \pubyear{2006}

\maketitle

\label{firstpage}

\begin{abstract}
Pulsar observations provide a suite of tests to
which stellar and binary evolutionary theory may compare. 
Importantly, the number of pulsar systems found from recent surveys 
has increased the statistical significance of pulsar population
synthesis results. 
To take advantage of this we are in the process of developing 
a complete pulsar population synthesis code 
that accounts for isolated and binary 
pulsar evolution, Galactic spatial evolution and pulsar survey 
selection effects. 
In a recent paper we described the first component of this code 
and explored how uncertainties in the parameters of binary and 
pulsar evolution affected the appearance of the pulsar population 
in terms of magnetic field and spin period. 
We now describe the second component which focusses on following 
the orbits of the pulsars within the Galactic potential. 
In combination with the first component we produce synthetic populations 
of pulsars within our Galaxy and calculate the resulting scale heights 
as well as the radial and space velocity distributions of the pulsars. 
Correlations between the binary and kinematic evolution of pulsars 
are also examined. 
Results are presented for isolated pulsars, 
binary pulsars and millisecond pulsars. 
We also test the robustness of the outcomes to variations in the assumed form 
of the Galactic potential, the birth distribution of binary positions, and the 
strength of the velocity kick given to neutron stars at birth. 
We find that isolated pulsars have a greater scale height than binary pulsars. 
This is also true when restricted to millisecond pulsars unless we allow 
for low-mass stars to be ablated by radiation from their pulsar companion 
in which case the isolated and binary scale heights are comparable. 
Double neutron stars are found to have a large variety 
of space velocities, in particular, some systems have 
speeds similar to the Sun.
We look in detail at the predicted Galactic population 
of millisecond pulsars with black hole companions, 
including their formation pathways, and show where 
the short-period systems reside in the Galaxy. 
Some of our population predictions are compared 
in a limited way to observations but the full potential 
of this aspect will be realised in the near future when 
we complete our population synthesis code with the 
selection effects component. 
\end{abstract}


\section{Introduction}
\label{s:intro}
 
Detailed examinations of the stellar populations within the 
Galaxy, and indeed the greater Universe, 
have unearthed many fascinating objects.
Examples of such systems are gamma-ray bursts 
(Klebesadel, Strong \& Olson 1973; Paczynski 1986; 
Bogomazov, Lipunov \& Tutukov 2008), coalescing double 
neutron stars (Rantsiou, Kobayashi, Laguna \& Rasio 2008), 
X-ray binaries (Schreier et al. 1972; 
Liu, van Paradijs \& van den Heuvel 2007; 
Galloway et al. 2008), microquasars (Margon et al. 1979;
Abell \& Margon 1979; 
Combi, Albacete-Colombo \& Marti 2008) and millisecond pulsars 
(Backer et al. 1982; Manchester et al. 2005).
Current belief suggests that all the aforementioned systems
arise from binary stars in which both stars are close 
enough to experience a strong gravitational interaction with their companion.
This may lead to mass transfer and depending
upon the the mass ratio, types and ages of the two stars a
plethora of different stellar and binary evolutionary phases may occur. 
Recent surveys spanning a large range of observed wavelengths 
now make it possible to monitor and analyse  
the combined properties of 
this variety of stellar populations.  
This in turn can place further constraints upon our theoretical 
understanding of stellar and binary evolution.
For example, with the quickly increasing number of known low-mass 
stellar X-ray binary systems ($187$: 
Liu, van Paradijs \& van den Heuvel 2007), 
high-mass X-ray binary systems ($114$; Liu, van Paradijs \& 
van den Heuvel 2006) and pulsars 
($> 1600$\footnote{http://www.atnf.csiro.au/research/pulsar/psrcat}; 
Manchester et al. 2005) 
it is possible to constrain features of 
neutron star (NS) and black hole (BH) formation from their 
relative Galactic scale heights.
These have been examined for NS and BH low-mass X-ray binaries
(via observations, Jonker \& Nelemans 2004) 
showing evidence that (contrary to previous belief) BHs may receive 
momentum from their formation mechanism -- the supernova 
(SN) event.

In this work our focus is on the Galactic population 
of pulsars. 
Observations of such objects 
occur primarily at radio (centimetre) wavelengths.
These systems have intrinsically weak signals (typically 
measured in mJy: Lorimer 2005) and are observed as 
a regular series of pulses in time.
Due to the frequency of the pulses we know these 
systems are compact (Hewish et al. 1968),
while the period derivative allows evolutionary models of 
these systems to be developed (e.g. Goldreich \& Julian 1969;
Ostriker \& Gunn 1969; Gunn \& Ostriker 1970;
Bisnovatyi-Kogan \& Komberg 1974; van den Heuvel 1984;
Kulkarni \& Narayan 1988; Chen \& Ruderman 1993) 
and tested (Dewey \& Cordes 1987; Tauris \& Bailes 1996; 
Dewi, Podsiadlowski \& Pols 2005; Kiel, Hurley, Bailes \& Murray 2008, 
hereafter Paper I).
Finally, distances may be estimated from the dispersion 
of the pulse due to the electron density distribution 
within the Galaxy (Cordes \& Lazio 2002).
From these observations we now believe that pulsars are
magnetic rotating neutron stars and that the radio signal 
arises from the magnetosphere and is well collimated.
Timing of these radio pulses gives the 
spin period $P$ and the spin period derivative 
$\dot{P}$ from which the magnetic field and characteristic
age of the pulsar may be inferred (see Paper I, and references 
therein for further details).

Knowledge of pulsar space velocities provides  
constraints on the effects of SNe on the NSs 
they give rise to.
This may, in-turn, place constraints on the SN 
mechanism and NS structure.
The birth velocities of pulsars (Lyne \& Lorimer, 1994) 
combined with the work of Gott, Gunn \& Ostriker (1970), 
Cordes, Romani \& Ludgren (1993), Dewey \& Cordes (1987) 
and Bailes (1989) demonstrated the necessity of 
asymmetric SN velocity kicks imparted on NSs by 
observing large isolated pulsar 
space velocities of order $1000$ km s$^{-1}$.
Later studies also found similar conclusions, again because 
observations suggest pulsar space velocities in excess of the mean for 
normal field stars (Fryer \& Kalogera, 2001). 
Evidence for asymmetric SNe is also provided by 
the misalignment of binary pulsar
spin vectors to the orbital angular momentum vector
(e.g. Kaspi et al. 1996).
The vectors would normally be expected to be coupled 
before the violent SNe because of tidal effects and 
any occurrence of mass transfer (Bhattacharya \& van den Huevel 1991;
Hurley, Tout \& Pols 2002).
There is observational evidence of this misalignment
for a number of pulsar binary systems including double
NS systems (e.g. Kramer 1998).

Another method in which we are able to constrain the effects of SNe
and to test the validity of our evolutionary assumptions 
(including the average magnitude of any kick delivered by a SN) 
is to compare, with observations, the kinematics of model pulsar 
populations within a model Galaxy (previous works include 
Dewey \& Cordes 1987, Bailes 1989, Lorimer et al. 1993, 
Lyne et al. 1998 and Sun \& Han 2004).
There are two obvious pulsar populations we may recognise, those that
are isolated pulsars and those pulsars within binary systems.
We may make further distinction with regards to the spin of a pulsar --
those of a `standard' spin period ($P > 1~$s), those of millisecond
spin periods ($P < 0.1~$s: known as millisecond pulsars, MSPs) while those 
between these two period ranges are known as partially spun-up or 
partially recycled pulsars.
The final pulsar population which we wish to point out here are those pulsars
which reside in double compact binaries, in particular double NS binaries
(such as PSR $1913+16$: Hulse \& Taylor 1975) and, within that population, 
double pulsar systems (PSR J$0737-3039$A\&B: Burgay et al. 2003; 
Lyne et al. 2004).

From a simple evolutionary analysis of these systems (see Paper~I) you may
expect there to be distinct scale heights for each pulsar population 
within the Galaxy.
This is assuming that all pulsars are given an asymmetric SN kick velocity
drawn from the same distribution (i.e. ignoring the possible
electron capture SN scenario which may confuse this issue: see Paper I;
Podsaidlowski et al. 2004; Ivanova et al. 2008).
Isolated pulsars -- it is reasonable to presume -- would have a greater scale 
height than those pulsars within binary systems.
This is because binaries are on average heavier than single stars
and also the binary orbit is able to absorb energy from the kick 
(as detailed in Hills 1983 and Tauris \& Takens 1998).
Having been previously ejected from a binary or having always 
been a single star would allow the kick velocity to have greater 
effect on the stellar space velocity.
Isolated pulsars that have felt two SN kicks during their
lifetime (one indirect and one direct) would be expected to 
have the largest scale height of any pulsar population (Bailes 1989).
Say, for instance, that the first of two massive stars disrupts 
a binary system, giving momentum to both stars out of the 
Galactic plane, then the second massive star (now isolated) 
undergoes a SN and receives further momentum.
Pulsars of different spin types, for example MSPs, 
can also be expected to have differing Galactic scale heights 
dependent upon the nature of their evolutionary histories. 
If one assumes a MSP forms via Roche-lobe overflow mass-transfer 
from a low mass
companion the system only passes through one SN event and 
this will occur in a close binary which can overcome larger
velocity kicks to stay bound than, say, a standard pulsar in a wider 
(pre-SN) orbit (e.g. Bailes 1989; Portegies Zwart \& Yungleson 1998).
If, however, some MSPs are formed via wind accretion from 
a high-mass companion (as discussed in Paper I), 
which may itself form a NS, a fraction of the MSP population 
may feel two kicks.
This may lead to an increase of the MSP Galactic scale height and 
also produce a method for isolated MSP production (as described by
Narayan, Piran \& Shemi 1991 and Paper I).
Finally, you may expect double pulsar systems 
to have a greater Galactic scale height than the single pulsar 
binary systems because they feel two SNe.
However, the fact that the binary system had to survive these two
kicks requires the kick velocity imparted from both events to
be small enough (or well directed) for the binary to survive.
Therefore, although two kicks occur it seems plausible that double pulsar
systems may not attain a greater scale height than their isolated cousins
(Pfahl, Podsiadlowski \& Rappaport 2005; Dewi, Podsiadlowski \& Pols 2005).
Further to this simple analysis, the relative 
ages of each system type will also play an important role 
in the observable scale height, because older systems
will have had more time to relax (outwards) in the Galactic gravitational
potential -- the distributions diffuse over time.
For a more in-depth review of pulsar evolution and kinematics 
see the living review of Lorimer (2008).

The work described in this paper allows us to address these issues and to 
predict and compare the Galactic spatial distributions of pulsar populations. 
We follow on from our pulsar population synthesis in Paper~I
to not only evolve the stellar evolution of pulsars but move 
them within the Galactic gravitational potential.
In other words we follow Galactic stellar, binary \textit{and} 
kinematic evolution. 
As introduced in Paper I we are developing a code comprised of
three modules: 
\textsc{binpop} (binary evolution), 
\textsc{binkin} (Galactic kinematics) and 
\textsc{binsfx} (synthetic survey simulations). 
An upcoming paper will describe the 
third module, \textsc{binsfx}, where we impose selection effects on the 
simulations, thus giving simulated data that can be 
compared directly to observations. 

Section 2 outlines our binary evolution code (\textsc{binpop}) and 
details a necessary update to follow 
the evolution of a system that is disrupted owing to an asymmetric SN 
velocity kick.
In Section~\ref{s:galaxy} we describe the Galaxy kinematic code 
(\textsc{binkin}) which integrates the positions of 
pulsars -- both in binary systems and isolated -- forward in time 
within the Galaxy. 
This includes details of how the initial conditions for the Galactic 
population are chosen and the parameters involved.
Results are given in Section~\ref{s:results} where, 
assuming a favoured binary and stellar evolutionary model of Paper~I, 
we examine the pulsar population scale heights and velocities 
that arise from different \textsc{binkin} model assumptions. 
This includes a detailed examination of the MSP-BH binary population. 
In particular, we explore the formation and evolution of these systems. 
This is followed by a discussion of our findings and the main uncertainties 
involved in Section~\ref{s:disc}.

\section{Rapid binary evolution}
\label{s:bkmodel}

The first module, \textsc{binpop}, was described in detail
in Paper~I.
Below we give an overview and also address the necessary 
modifications to \textsc{binpop} in order to correctly follow 
the Galactic positions of both members of a disrupted binary system. 

\subsection{BINPOP}
\label{s:binpop}
\textsc{binpop} is a stellar/binary population synthesis package
which convolves the binary stellar evolution (\textsc{bse}) 
code of Hurley, Tout \& Pols (2002: hereafter HTP02) 
with realistic initial stellar and binary parameter 
distributions (as developed in Kiel \& Hurley 2006 and Paper~I).
Stellar evolution is included according to the formulae presented
in Hurley, Pols \& Tout (2000).
Meanwhile, \textsc{bse} attempts to account for all important 
binary evolutionary processes.
These include tidal evolution, mass transfer, common envelope 
(CE) evolution, stellar mergers, magnetic braking, orbital 
gravitational radiation and supernovae velocity kicks.
In Paper~I extensive additions were made to \textsc{bse},
in terms of NS physics, so that pulsar evolution could be 
followed in detail.
This means that aspects such as  magnetic field decay, accretion 
induced field decay and spin-up, propeller evolution and pulsar
death lines are now included.
Inherent uncertainties in the variety of binary and pulsar evolutionary 
processes requires a host of parameterised prescriptions to be 
incorporated into \textsc{bse}.
For example our lack of understanding of CE evolution is 
expressed as a parameter, $\alpha_{\rm CE}$, often referred to as
the efficiency parameter.
In terms of pulsar evolution there is the magnetic 
field decay time-scale, $\tau_{\rm B}$ and the accretion induced
decay time-scale, $k$, for example, which are uncertain.
Over time the uncertainty in many parameters has decreased 
-- albeit slightly --
due to a large array of simulations and increasingly detailed 
observations (e.g. Lyne et al. 1998; Portegies Zwart \& Yungelson 1998; 
HTP02; Belczynski, Kalogera \& Bulik 2002; 
Podsiadlowski, Rappaport \& Han 2003; 
Sun \& Han 2004; Yusifov \& Kucuk 2004; Hobbs, Lorimer, Lyne \& Kramer 2005; 
Kiel \& Hurley 2006; Cordes et al. 2006; Lorimer et al. 2006; 
Ferrario \& Wickramasinghe 2007; 
Liu, van Paradijs \& van den Heuvel 2007; 
O'Shaughnessy, Kim, Kalogera \& Belczynski 2008;
Belczynski et al. 2008 and Paper~I).

\subsection{Binary evolution SN kick update}
\label{s:binupdate}
When following the kinematic evolution of a binary system within 
the Galaxy we require knowledge of 
the Galactic gravitational potential
-- the acceleration felt on the binary centre of mass (CofM) 
owing to the Galaxy -- 
as well as any internal sources of momentum that arise.
The primary stellar evolutionary phase that can 
perturb an orbit or disrupt a binary system is a SN.
If the SN occurs in a binary and enough material 
is ejected from the system during the event (more than half
of the total binary mass) the binary may disrupt (Hills 1983).
Along with the assumed instantaneous mass loss, if there is  
any asymmetry in the explosion (which is arguable in 
the case of BHs, see Podsiadlowski, Rappaport \& Han 2003
but note Pfahl, Podsiadlowski \& Rappaport 2005), 
the newly formed compact star will receive a velocity kick which 
the binary CofM will feel 
(Shklovskii 1970; Lyne \& Lormier 1994; Tauris \& Takens 1998; 
HTP02).
Depending upon the velocity kick direction and magnitude this 
may disrupt a binary or save it from mass loss-induced dissipation 
(Hills 1983; Kalogera 1998; Pfahl, Rappaport \& Podsaidlowski 2003).

The algorithm described by HTP02 allows realistic orbital evolution 
modelling if the binary system survives the blast
and stays gravitationally bound (eccentricity, $e < 1$).
However, because binary systems cease to exist once they become
unbound, HTP02 were not troubled with calculating the
recoil escape velocities of the two disassociated stars
traveling on hyperbolic orbits with $e > 1$.
Now that the Galactic spatial kinematics of both binary systems and 
isolated stars is a concern, knowledge of all associated 
velocity changes are required.
To this end we formulate a disruption model (in Section~\ref{s:simple})
and also describe similar methods derived by other groups 
(in Section~\ref{s:TNT}).
All methods considered here are generalised to allow for initially
eccentric systems.

\begin{figure}
  \includegraphics[width=84mm]{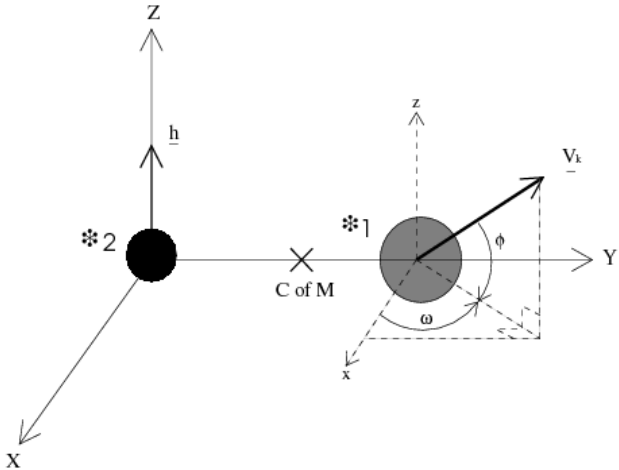}
  \caption{
  HTP02 orbital geometry as asymmetric SN occurs.
  Taken from Figure A1 of HTP02.
   \label{f:fig1}}
\end{figure}

\subsubsection{BSE disruption model}
\label{s:simple}
\begin{figure}
  \includegraphics[width=84mm]{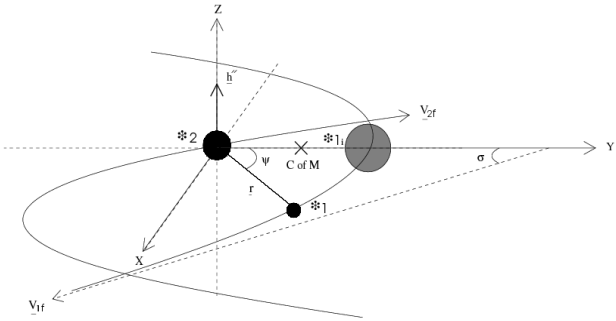}
  \caption{
  Orbital geometry of our disruption method
  after an asymmetric SNe.
   \label{f:fig2}}
\end{figure}

Before one can calculate the final velocities of the disrupted
stars the binary system must be known to disrupt.
We start by considering what is already within the capabilities of the 
rapid binary evolution code (cf. Appendix A of HTP02) which we
outline here.
This assumes a reference frame in-which the pre-SN CofM is at rest, 
$\textbf{V}_{\rm s} = 0$, 
and the secondary star (the star not exploding) is at the origin
(as shown in Figure~\ref{f:fig1}).
The magnitude of the pre-SN relative orbital velocity is
\begin{equation}
  \label{e:vorb}
V_{\rm orb} = \sqrt{\mu \left( \frac{2}{r} - \frac{1}{a} \right)}
\end{equation}
and vectorily is
\begin{equation}
\label{e:relvel}
\textbf{V} = -V_{\rm orb}\left( \sin{\beta}\hat{\textbf{x}} + \cos{\beta}\hat{\textbf{y}} \right).
\end{equation}
The separation of the two stars is $\textbf{r} = r[0,~1,~0]$ 
and $\beta$ is the angle between $\textbf{r}$ and $\textbf{V}$. 
Also, $\mu = GM_{\rm b}$ where $G$ is the gravitational constant,
$M_{\rm b} = M_1 + M_2$ and the subscripts denote the particular star 
(the primary star being $_1$ and the secondary $_2$).
In this reference frame the two pre-SN stellar velocities are,
\begin{equation}
  \label{e:eq7}
\textbf{V}_1 = \frac{M_2}{M_{\rm b}}\textbf{V}
\end{equation}
and
\begin{equation}
\textbf{V}_2 = -\frac{M_1}{M_{\rm b}}\textbf{V}.
\end{equation}
Furthermore, the orbital angular momentum $\textbf{J}$ is expressed as,
\begin{equation}
  \label{e:orbJ}
\textbf{J} = \frac{M_1M_2}{M_{\rm b}}\textbf{r} \times \textbf{V} = \frac{M_1M_2}{M_{\rm b}}\textbf{h} = M_1M_2\sqrt{\frac{l}{\mu}},
\end{equation}
where  $\textbf{h}$ is the  specific angular momentum (aligned with 
the z-axis) and $l = a\left(1 - e^2 \right)$ is the semi-latus rectum.
The primary star, the SN progenitor, is about to explode 
and receive a momentum impulse arising from a velocity kick,
\begin{equation}
\textbf{V}_{\rm kick} = V_{\rm kick}\left( \cos{\omega}\cos{\phi}\hat{\textbf{x}} + \sin{\omega}\cos{\phi}\hat{\textbf{y}} + \sin{\phi}\hat{\textbf{z}} \right)
\end{equation}
(where $\hat{\textbf{x}}$, $\hat{\textbf{y}}$ and $\hat{\textbf{z}}$ 
are the unit directions vectors).
The kick speed, $V_{\rm kick}$, is modelled by a Maxwellian 
distribution,
\begin{equation}
\label{e:max}
P \left(V_{\rm kick} \right) = 
     \sqrt{\frac{2}{\pi}} \frac{V_{\rm kick}^2}{V_\sigma^3} \exp^{-\frac{V_{\rm kick}^2}{2V_\sigma^2}},
\end{equation}
as given by Hansen \& Phinney (1997) with a dispersion $V_\sigma$.
To facilitate any possible comparison to previous works, such as
Hansen \& Phinney (1997), Portegies Zwart \& Yungelson (1998),
HTP02 or Paper I, we take $V_\sigma = 190~$km s$^{-1}$.
However, a value of $265~$km s$^{-1}$ has more recently been 
suggested by Hobbs et al. (2005) and there have also been 
suggestions of a bimodal kick distribution 
(Arzoumanian, Chernoff \& Cordes 2002).
The direction of the kick is specified by chosing two angles 
$\omega$ and $\phi$ within the ranges 
$0 < \omega < 2\pi$ and $-\pi/2 < \phi < \pi/2$ (as shown in 
Figure~\ref{f:fig1}).

Immediately after the asymmetric SN event the new velocity of the
proto-NS is 
\begin{equation}
\textbf{V}^{'}_1 = \textbf{V}_1 + \textbf{V}_{\rm kick}
\end{equation}
and taking into account the instantaneous SN mass loss $\Delta M$ 
from the system we have a new relative velocity, 
\begin{equation}
\textbf{V}^{'} = \textbf{V} + \textbf{V}_{\rm kick}.
\end{equation}
The velocity of the new centre of mass relative to the old 
centre of mass is
\begin{equation}
\label{e:vs}
\textbf{V}^{'}_{\rm s} = \frac{M_{\rm NS}}{M^{'}_{\rm b}} \textbf{V}_{\rm kick} + \frac{\Delta M M_2}{M^{'}_{\rm b} M_{\rm b}}\textbf{V}
\end{equation}
(as in HTP02 equation A14).
Following HTP02 the new eccentricity, $e^{'}$, and semi-major 
axis, $a^{'}$, of the system can then be calculated.
If this gives an eccentricity greater than unity then the 
system is disrupted and we need expressions for the runaway 
velocities of the stars.
For further details on the SN treatment and the evolution 
of binary systems which survive the event see HTP02. 

Prior to the SN our chosen coordinate system had the
orbital angular momentum vector directed along the $z$-axis
but this will no longer be true for the post-SN system
(unless $\textbf{V}_{\rm kick} = 0$).
However to simplify our post-SN calculations it is desirable 
to realign the vector with the $z$-axis.
This realignment in the $x$-$z$ plane, owing to the SN, 
is performed by a rotation, $R_{\rm xz}$, around the y-axis 
such that the post-SN unaligned orbital specific angular momentum, 
$\textbf{h}^{'} = \textbf{r} \times \textbf{V}^{'} 
= [0,~r,~0] \times [V_{\rm x},~V_{\rm y},~V_{\rm z}]$ becomes
$\textbf{h}^{''} = [0,~0,~h^{''}_{\rm z}]$.
Note that we are using $^{'}$ to denote the frame immediately 
after the SN and $^{''}$ when referring to the coordinate 
system of the frame after rotation.
Here the rotation is guided by $\nu$, the angle between 
the pre- and post-SN angular momentum vectors (see Equation A13 of HTP02).
This rotation matrix allows us to map our final velocities back onto
the original coordinate system.

Now we consider the post-SN motion of the two stars which is 
governed by a hyperbolic conic section.
We can no longer assume that the separation vector between 
the stars is aligned along the y-axis because the 
system is not necessarily at periastron.
To account for this possible shift in coordinate system
around the z-axis we calculate where in the orbit each star resides.
Here we have
\begin{equation}
\cos{\psi} = \frac{1}{e^{'}}\left( \frac{l^{'}}{r} - 1\right) = \frac{1}{e^{'}}\left( \frac{(h^{'})^2}{GM_{\rm b}^{'}r} - 1\right)
\end{equation}
with $\psi$ defined to be positive in the positive
y-direction and zero along the positive y-axis.
There are two possible hyperbolic orbits for which
each star may travel along -- one in the positive y-region, the other 
in the negative y-region -- which is governed by the direction
of the y-component of the new relative velocity, that is,
$\textbf{r} \cdot V^{'}$.
The sign of $\psi$ depends upon the the sign of the 
$\textbf{r} \cdot V^{'}$ value, where 
$\psi < 0$ when $\textbf{r} \cdot V^{'} > 0$.
The post-SN binary mass must also be updated:
$M^{'}_{\rm b} = M_{\rm NS} + M_2$ with $M_{\rm NS}$ being 
the primary star mass.
Using this and the new semi-major axis we may now calculate the final 
velocities of the two stars in the pre-SN centre of mass reference frame. 
Assuming a velocity at infinity, 
$\textbf{V}_\infty$, which is directed along an asymptote of the 
hyperbolic orbit we have final velocities for the two stars of
\begin{equation}
  \label{e:v1f}
\textbf{V}_{\rm 1f} =  \textbf{V}^{'}_{\rm s} - R_{\rm xz}^{-1}\frac{\rm M_2}{M^{'}_{\rm b}}\textbf{V}_\infty
\end{equation}
and
\begin{equation}
  \label{e:v2f}
\textbf{V}_{\rm 2f} = \textbf{V}^{'}_{\rm s} + R_{\rm xz}^{-1}\frac{M_{\rm NS}}{M^{'}_{\rm b}} \textbf{V}_\infty.
\end{equation}
A simple calculation gives the magnitude,
\begin{equation}
V_\infty = \sqrt{\frac{G M^{'}_{\rm b}}{a^{'}}}.
\end{equation}
The angle of the hyperbolic asymptote may be calculated 
from the angle $\sigma$ (as shown from Figure~\ref{f:fig2} 
which describes the post-SN coordinate system) where
$\cos{\sigma} = 1/e^{'}$ (as in Tauris \& Takens 1998) 
and $\sigma$ is always positive.
This restricts $\sigma$ to range from $0 \rightarrow \pi/2$.
With our two angles we define the difference angle 
$\gamma = \sigma - \psi$ which is used in 
rotating the coordinate system around the z-axis.
Separating $\textbf{V}_{\rm 1f}$ and $\textbf{V}_{\rm 2f}$ 
into component form gives us:
\begin{eqnarray}
\label{e:Vkhi}
V_{\rm 1fx} &=& \frac{M_{\rm NS}}{M^{'}_{\rm b}}V_{\rm kick}\cos{\omega}\cos{\phi} + \frac{\Delta M M_2}{M^{'}_{\rm b} M_{\rm b}} V_{\rm orb}\sin{\beta} \nonumber \\
&& - V_\infty \cos{\nu} \sin{\gamma},
\end{eqnarray}
\begin{eqnarray}
V_{\rm 1fy} &=& \frac{M_{\rm NS}}{M^{'}_{\rm b}}V_{\rm kick}\sin{\omega}\cos{\phi} + \frac{\Delta M M_2}{M^{'}_{\rm b} M_{\rm b}} V_{\rm orb}\cos{\beta} \nonumber \\
&& - V_\infty \cos{\gamma}, 
\end{eqnarray}
\begin{eqnarray}
V_{\rm 1fz} &=& \frac{M_{\rm NS}}{M^{'}_{\rm b}}V_{\rm kick}\sin{\phi},
\end{eqnarray}

\begin{eqnarray}
V_{\rm 2fx} &=& \frac{M_{\rm NS}}{M^{'}_{\rm b}}V_{\rm kick}\cos{\omega}\cos{\phi} + \frac{\Delta M M_2}{M^{'}_{\rm b} M_{\rm b}} V_{\rm orb}\sin{\beta} \nonumber \\
&&  + V_\infty \cos{\nu} \sin{\gamma}, 
\end{eqnarray}

\begin{eqnarray}
V_{\rm 2fy} &=& \frac{M_{\rm NS}}{M^{'}_{\rm b}}V_{\rm kick}\sin{\omega}\cos{\phi} + \frac{\Delta M M_2}{M^{'}_{\rm b} M_{\rm b}} V_{\rm orb}\cos{\beta} \nonumber \\
&&  + V_\infty \cos{\gamma} 
\end{eqnarray}
and
\begin{eqnarray}
\label{e:Vkhf}
V_{\rm 2fz} &=& \frac{M_{\rm NS}}{M^{'}_{\rm b}}V_{\rm kick}\sin{\phi}.
\end{eqnarray}

We also account for coalescence of the two stars if the newly formed
compact star velocity kick is directed towards the companion.
Coalescence occurs if the companion radius, $R_2$, is greater than 
periastron or the distance of closest approach: if
$R_2 > a^{'}(e^{'} - 1)$ we assume the stars 
coalesce (see also Tauris \& Takens 1998; Belczynski et al. 2008) 
and the merger 
outcome has the final centre of mass velocity, $\textbf{V}^{'}_{\rm s}$
and the mass of the NS and companion are combined (see HTP02
for further details of merger outcomes).

Equations~\ref{e:Vkhi} to \ref{e:Vkhf} are the velocities calculated 
within the \textsc{bse} kick routine and are communicated 
into \textsc{binkin}.
Before adding these to the Galactic binary 
centre of mass velocity at the time of SN it is first necessary
to perform a random orientation of the pre-SN binary orbital
plane, which until now has been fixed in the Galactic xy-plane.
We randomly choose three Euler angles $\alpha_{\rm E}$, 
$\beta_{\rm E}$ and $\gamma_{\rm E}$, within the ranges
$0 \leq \alpha_{\rm E}{\rm ,\gamma_{\rm E}<2\pi}$ and 
$0 \leq \beta_{\rm E} < \pi$, to give a 3D rotation
of the velocities in Equations~\ref{e:v1f} and \ref{e:v2f}.
The post-SN velocities (pre-SN CofM velocity plus the 
rotated disruption velocities) are then used within the 
kinematic routine to calculate the subsequent velocities 
and positions within the Galaxy of the pulsar and its 
former companion.

\subsubsection{Related disruption models}
\label{s:TNT}

There are two other groups who have independently developed
models for deriving the run-away velocities of stars
from a disrupted binary.
The method published recently by 
Belczynski et al. (2008)
is similar to our demonstration and is also generalised
for arbitrary eccentricity.
The main difference is that it allows a velocity
$\textbf{v}_{\rm imp}$ to be imparted to the secondary
from the expanding shell of the SN.
In our work we essentially assume that $\textbf{v}_{\rm imp} = 0$
which has minimal effect except in some cases of small pre-SN 
orbital separation (Kalogera 1996; Belczysnki et al. 2008).
Tauris \& Takens (1998; hereafter T\&T98) have developed
a relatively sophisticated model.
The major difference between our model and the T\&T98 model is the 
coordinate system scheme and the latters assumption that the companion 
star may have momentum imparted directly onto it 
from the SN blast wave (as in Belczynski et al 2008 and 
Dewey \& Cordes 1987).
The companion star may also have some fraction of mass stripped
off it and/or ablated owing to the impact of the shell of material
ejected from the primary.
To include the possibility of investigating the effect of these
additional considerations we have worked through the
T\&T98 demonstration and implemented it as an option in \textsc{bse},
generalised to eccentric orbits.
However, we do not exercise this option in this work.

\subsubsection{Disruption model illustration}
\label{s:modelcomp}

To illustrate the effect the binary orbit has on the runaway 
velocities of disrupted stars we produce a simple population 
of binary systems.
For this population the primary mass, $M_1$, is randomly 
selected from a flat distribution ranging from 
$10 - 20~M_\odot$, the secondary mass, $M_2$, from a flat 
distribution ranging from $0.1 - 20~M_\odot$ 
and the orbital separation is selected randomly from a flat 
distribution from $1$ to $10~000~R_\odot$.
All systems are initially circular to simplify 
the analysis.
The radius of the secondary star is linked to its mass by
$R_2 = 1.3M_2^{0.6}$ if $M_2$ is greater than unity 
and by $R_2 = M_2$ otherwise.
We make sure the system is not in contact at birth.
We then let the primary undergo a SN that leaves a NS
with $M_{NS} = 1.4~M_\odot$ and assume the remnant is
given a kick from Equation~\ref{e:max} with $V_\sigma = 190~$km s$^{-1}$
(in accordance with Hansen \& Phinney 1997).
The post-SN velocities for disrupted stars are calculated using
the \textsc{bse} method detailed above.

\begin{figure}
  \includegraphics[width=84mm]{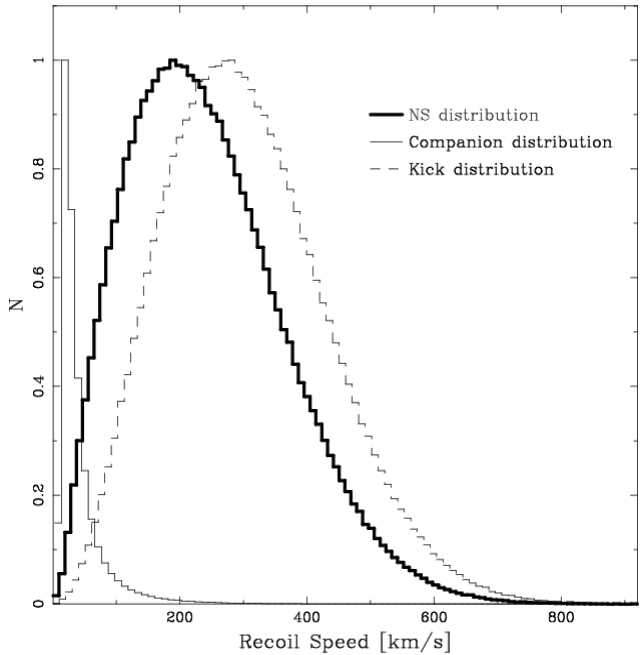}
  \caption{
  The speeds of the two stars following the SNe.
  The thick line represents the NS recoil speed
  distribution while the thin line is the companion
  recoil speed distribution.
  Included is the assumed asymmetric SN kick distribution
  assuming a dispersion of $190~$km s$^{-1}$ (dashed line).
   \label{f:fig3}}
\end{figure}

After generating a million systems we find that the majority
($99\%$) become unbound and the incidence of coalescence is 
negligible.
In Figure~\ref{f:fig3} we see the distributions of
NS and companion star recoil speed and compare this to
the $\textbf{V}_{\rm kick}$ distribution, i.e. the
distribution for a population of standard single NSs.
The first item we wish to note is the difference in the 
typical velocities received by both stars: the NS, which directly
experiences the additional momentum imparted from the asymmetric
SN, will most likely depart the binary system 
with a greater velocity than the companion star (relative to the
CofM).
The second point of interest is the similarity between the 
NS recoil speed distribution and the kick distribution.
Clearly not all of the momentum imparted onto the NS goes into
the NS recoil velocity, some of the momentum is instead transported into 
the CofM momentum, consumed by the disruption of the binary system
and converted into additional velocity of the companion star.
Therefore, although the shape of the NS recoil speed distribution is 
consistent with the kick distribution the NS distribution is shifted
somewhat to lower values.
In regards to this, observational pulsar velocity studies 
that do not account for the possibility of a fraction of the 
sample being disrupted from binary systems may underestimate 
the underlying SN kick distribution.
It also suggests a possible mechanism for any bimodality found
in the velocity structure of pulsar observations --  
similar to that found by Arzoumanian, Chernoff \& Cordes (2002) 
from which they concluded a bimodal asymmetric SN kick 
distribution, or possibly binary disruption effects, could 
cause such detected velocities. 
We note that the form of the underlying orbital period 
(or separation) distribution of the model binaries will 
affect the distribution of recoil speeds 
-- with an increased proportion of short-period systems leading to an 
increased difference between the NS recoil speed and kick distributions 
-- and we have not explored this aspect in detail here.

\section{Galactic kinematics}
\label{s:galaxy}

\subsection{Galactic gravitational potentials}
\label{s:potentials}

Much work over the years has lead to estimates of 
the Galactic gravitational potential.
Miyamoto \& Nagai (1975) generalised the work of Toomre (1963)
who calculated flattened Plummer (1911) models for the Galaxy.
Since then further observations have lead to estimates by
Carlberg \& Innanen (1987) of disk-halo, bulge and nucleus 
potentials which in turn have been updated by Kuijken \& Gilmore (1989).
Kuijken \& Gilmore (1989) used more extensive
observations of Galactic stellar densities, which
allows the mapping to an assumed (to first order) smooth 
time-independent galactic gravitational potential.

The KG89 model potential is,
\begin{eqnarray}
\label{e:KGpot}
\Phi_{\smfont{G}}^{KG} &=& \Phi_{\smfont{disk/halo}}^{KG} + \Phi_{\smfont{nuc}}^{KG} + \Phi_{\smfont{bulge}}^{KG} \\
\end{eqnarray}
where,
\begin{eqnarray}
\Phi_{\smfont{disk/halo}}^{KG} &=& -\frac{GM_\smfont{disk}}{\sqrt{\left(a + \sum_{\rm i=1}^3\beta_i\sqrt{z^2+h_{\rm i}^2} \right)^2 + b^2 + r^2}} \\
\Phi_{\smfont{nuc}}^{KG} &=& -\frac{GM_\smfont{nuc}}{\sqrt{b^2 + r^2}} \\
\Phi_{\smfont{bulge}}^{KG} &=& -\frac{GM_\smfont{bulge}}{\sqrt{b^2 + r^2}},
\end{eqnarray}
and the parameter values for each region are:
\begin{description}
\item [\textbf{disk/halo:}] $\beta_1 = 0.4$, $\beta_2 = 0.5$, $\beta_3 = 0.1$, $h_1 = 0.325~{\rm kpc}$, $h_2 = 0.090~{\rm kpc}$, $h_3 = 0.125~{\rm kpc}$, $a = 2.4~{\rm kpc}$, $b = 5.5~{\rm kpc}$, $M_{\rm disk} = 1.45\times10^{11}~M_\odot$
\item [\textbf{nucleus:}] $b = 0.25~{\rm kpc}$, $M_{\rm nuc} = 9.3\times10^9~M_\odot$
\item [\textbf{bulge:}] $b = 1.5~{\rm kpc}$, $M_{\rm bulge} = 1.0\times10^{10}~M_\odot$.
\end{description}

If necessary, use of the KG89 model allows us to compare results to 
previous binary pulsar population synthesis works such 
as Lorimer et al. (1993).
While now considered somewhat outdated, the KG89 potential
is still in use within recent observational works, such as, 
Freire, Ransom \& Gupta (2007) who use it in their calculation 
of the pulsar spin period when accounting 
for observational effects of the acceleration between the 
Solar System Barycentre and NGC 1851 - the globular cluster 
in which their observed pulsar resides.
However, a recent appraisal of the most promising Galactic gravitational 
potentials to use was completed by Sun \& Han (2004).
Their favoured method for ease of implementation is that given by 
Paczynski (1990: see below).
Sun \& Han (2004) also commented favourably on the work of 
Dehnen \& Binney (1998) who fit a multi-parameter mass model to 
kinematic data of the Milky Way.
Sun \& Han (2004) found  that the Dehnen \& Binney (1998) model
is overly complicated to set up and manipulate,
while the simpler Paczynski (1990, hereafter P90) model is as accurate as the
Dehnen \& Binney (1998) model.
As such we also include the P90 model in our work.

The P90 model, like the KG89 model,
follows the potential of Miyamoto \& Nagai (1975)
for the disk ($i = 1$) and spheroid components ($i = 2$).
Equation 9 of Paczynski (1990) is,
\begin{equation}
\label{e:Pac1}
\Phi_i^{\rm P}(R,z) = -\frac{GM_i}{\left( R^2 + \left[ a_i+(z^2 + b_i^2)^{(1/2)}\right]^2\right)^{(1/2)}},
\end{equation}
where $R = \sqrt(x^2 + y^2)$.
The P90 model, however, differs from the KG89 model 
not only with the assumed constant values used within 
Equation~\ref{e:Pac1} (for the disk potential) but also 
with the assumed form of the halo potential,
\begin{equation}
\label{e:Pac2}
\Phi_{\rm h}^{\rm P}(r) = -\frac{GM_{\rm h}}{r_{\rm h}}\left[ \frac{1}{2} \ln(1+\frac{r^2}{r_{\rm h}^2}) + \frac{r_{\rm h}}{r}~{\rm atan}\left(\frac{r}{r_{\rm h}}\right) \right],
\end{equation}
where $r^2 = R^2 + z^2$ is used to simplify the equation.
The parameters being:
\begin{description}
\item[\textbf{disk (i = 1):}] $a_1 = 0~{\rm kpc},~b_1 = 0.277~{\rm kpc},~M_1 = 1.12\times10^{10}~M_\odot$,
\item[\textbf{spheroid (i = 2):}] $a_2 = 3.7~{\rm kpc},~b_2 = 0.20~{\rm kpc},~M_2 = 8.07\times10^{10}~M_\odot$,
\item[\textbf{halo (i = h):}] $r_{\rm h} = 6.0~{\rm kpc},~M_{\rm h} = 5.0\times10^{10}~M_\odot$.
\end{description}

The KG89 and P90 models are both based on 
old observations of the stellar neighbourhood.
As such, these models are only considered accurate 
out to a radii of $\sim 12$ kpc.
More recent observations completed in the Sloan 
Digital Sky Survey (SDSS: York et al. 2000), 
within the Sloan Extension for Galactic Understanding 
and Exploration (SEGUE: Lee et al. 2008) program,
have allowed the Galactic gravitational potential 
to be measured out to a radii of $\sim 60$ kpc 
(Xue et al. 2008).
To do this Xue et al. (2008) have made line of 
sight velocity measurements of $\sim 2500$ blue 
horizontal branch stars which are converted into 
circular velocity estimates of the Milky Way.
Ultimately the work of Xue et al. (2008) is completed 
to probe the halo of our Galaxy and thus they do not 
examine in any detail the inner Galactic potential.
However, at this stage we consider their complete 
assumed Galactic gravitational potential as an option 
in our work.
The Xue et al. (2008, hereafter Xue08) model makes use of the 
following exponential disk, Hernquist (1990) bulge 
and Navarro, Frenk \& White (1996; NFW) halo potentials 
respectively:
\begin{equation}
\label{e:Xue1}
\Phi_{\rm disk}^{\rm X}(r) = -\frac{GM_{\rm disk}^{\rm X}}{r}\left[ 1-\exp^{-r/b}\right],
\end{equation}
\begin{equation}
\Phi_{\rm bulge}^{\rm X}(r) = -\frac{GM_{\rm bulge}^{\rm X}}{r+C_{0}}
\end{equation}
and
\begin{equation}
\label{e:Xue3}
\Phi_{\rm NFW}^{\rm X}(r) = \frac{GM_{\rm vir}}{rC_{\rm x}^{'}}\ln \left( 1 + \frac{C_{\rm x}}{r_{\rm vir}}r \right).
\end{equation}
Here $C_{\rm x}^{'} = \ln(1+C_{\rm x}) - C_{\rm x}/(1+C_{\rm x})$,
while the values used within Xue08 vary depending upon
the assumed halo description.
We also note here that the form of $\Phi_{\rm NFW}^{\rm X}(r)$ is exactly
that of Smith et al. (2007), who provide differing values for the 
virial mass $M_{\rm vir}$, radius $r_{\rm vir}$ and concentration $C_{\rm x}$.
Xue08 match their observed circular Galactic stellar velocity 
estimates to smooth particle hydrodynamical simulations from which 
they provide values for $M_{\rm vir}$, $r_{\rm vir}$ and $C_{\rm x}$.
The values from Xue08 used in our work are:
\begin{description}
\item[\textbf{disk:}] $M_{\rm disk}^{\rm X} = 5\times10^{10}~M_\odot$, $b = 4~{\rm kpc}$,
\item[\textbf{bulge:}] $M_{\rm bulge}^{\rm X} = 1.5\times10^{10}~M_\odot$, $C_0 = 0.6~{\rm kpc}$,
\item[\textbf{NFW:}] $M_{\rm vir}^{\rm X} =1.03\times10^{12}~M_\odot$, $r_{\rm vir} = 278~{\rm kpc}$, $C_{\rm x} = 11.8$.
\end{description}

These Galactic gravitational potential models are now a part 
of \textsc{binkin}, updating the original algorithm based on
Lorimer, Bailes \& Harrison (1997), and extending upon similar
population synthesis models such as that of Faucher-Giguere \& Kaspi (2006).

\begin{figure}
  \includegraphics[width=84mm]{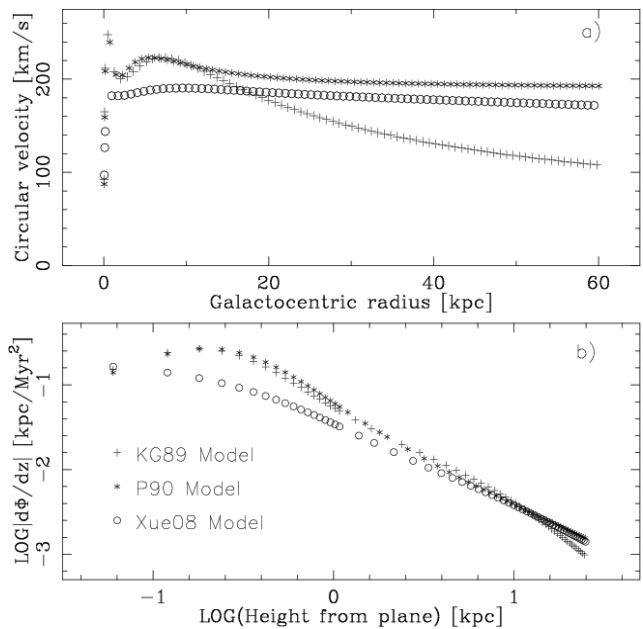}
  \caption{
  The three Galactic gravitational models are depicted
  in two different manners here to illustrate their properties and
  differences.
  The top panel (a) shows the circular velocity for a 
  range of Galactocentric radial positions in the plane
  of the Galactic potential.
  The bottom panel (b) depicts modulus acceleration 
  in increasing height from the plane.
   \label{f:fig4}}
\end{figure}

Figure~\ref{f:fig4} depicts the three assumed 
Galactic gravitational models.
In particular we wish to point out the inner region 
of the Xue08 model, which contains a smaller restoring 
force than the other two models.
We return to this later in Section~\ref{s:results}.
The KG89 gravitational potential model decays 
faster than the other two models beyond the central 
Galactic region and the P90 model rotational curve 
follows the KG89 model in the inner region
of the Galaxy while beyond a Galactic radius of $\sim 20~$kpc
the rotational curve flattens off similarly to the Xue08 model.

\subsection{Initial conditions and integration method}
\label{s:method}

Once we have our assumed Galactic potential, which
is in cylindrical coordinates, $\Phi\left( r,\phi, z\right)$,
the progenitor pulsar systems must be given some initial Galactic
position, $R_{\rm init}$, selected randomly from a given distribution.
The most straight forward distribution is to assume 
a thin disk with some maximum height and radius.
This simple method allows the system to relax over 
time into a similar distribution to the observed stellar 
number density distribution in height with respect
to the plane (see distributions given in Sun \& Han 2004).
However, as shown in Section~\ref{s:initdist} this 
overestimates the number of systems 
(in this case pulsars) found in the central region -- 
the deficiency of observed pulsars within the central 
region is believed to be a real lack of pulsars and not 
caused solely by observation selection effects 
(Lorimer et al. 2006).
However, Lorimer et al. (2006) caution readers that 
more observations are required to provide a definitive result.

To combat this over density within the central regions 
a preferred option is to use the distribution of SN remnants
developed in Paczynski (1990): 
\begin{equation}
\label{e:PacRdist}
P(R_{\rm init})dR = a_{\rm R}\left(R_{\rm init}/R^2_{\rm exp}\right) \exp{\left(-R_{\rm init}/R_{\rm exp}\right)}dR \, ,
\end{equation}
where $R_{\rm exp} = 4.5$ is simply an exponential scale
length of the radial distribution and $a_{\rm R}$ is a constant 
of integration equal to $1.0683$ over $R = 0 \rightarrow 20~$kpc.
A third option is to use the distribution derived by 
Yusifov \& Kucuk (2004; their Equation 17) from observations of OB type 
Population I stars:
\begin{equation}
\label{e:YKRdist}
P(R_{\rm init})dR = a_{\rm R}\left(R_{\rm init}/R_\odot \right)^4 \exp{\left(-b\frac{R_{\rm init}}{R_\odot}\right)}dR.
\end{equation}
If OB stars are assumed to be the progenitors of NSs then
this can be taken as the Galactic pulsar progenitor birth radial 
distribution.
Here $a_{\rm R} \sim b^5/24 \sim 606$ 
(for $0 \geq R_{\rm init} \geq 20~$kpc).
We utilise all three radial distributions -- thin disk, 
Paczynski (1990) and Yusifov \& Kucuk (2004) -- in our 
models to generate birth locations.
However, because the Paczynski (1990) distribution has 
had much use in the past we assume this to be our 
standard description.

In terms of the initial distribution of systems in height
from the plane we simply assume a uniform distribution with
maximum height of $|z_{\rm maxi}|$.
Systems over time relax outwards in $|z|$ and such a simple
initial $|z|$ distribution compares well to those favoured
in Sun \& Han (2004).
Furthermore, according to Paczynski (1990) as long as the 
systems are born relatively close to the Galactic plane (few hundred
parsec) the initial distribution in $|z|$ for energetic populations 
is redundant.
In the future the initial spatial distribution will contain 
spiral arms, similar to that completed within Faucher-Giguere 
\& Kaspi (2006) who suggest that because Galactic arm structure
is visible in large observational surveys it is necessary 
for any realistic pulsar population synthesis simulation to model 
this structure.

Once a position is found for a system an initial velocity
is simply calculated from the (estimated)
Galactic gravitational potential at that point.
With knowledge of the position and space velocity the next 
step is to solve four coupled equations of motion to evolve
the system position forward in time.
These are (Paczynski 1990):
\begin{eqnarray}
&& \frac{dR}{dt} = V_{\rm R},~\frac{dz}{dt} = V_{\rm z},~\frac{dV_{\rm R}}{dt} = -\frac{\partial \Phi}{\partial R} + \frac{L^2}{R^3}~{\rm and}~ \\
&& \frac{dV_{\rm z}}{dt} = -\frac{\partial \Phi}{\partial z}. \nonumber
\end{eqnarray}
These four equations are found by calculating the acceleration 
in $R$, $\phi$ and $z$ ($V_{\rm R}~{\rm and}~V_{\rm z}$ are velocities
in $R$ and $z$ respectively) induced onto a test particle by the 
gravitational potential. Assuming an axisymmetric potential 
around $z$ produces constanst angular momentum, $L$, felt by 
a test particle.
To integrate forward in time a fourth order Runge-Kutta
integration routine is used, similar to that used by Paczynski (1990)
and Lorimer et al (1993).

It is now possible for us to evolve the complete Galactic 
orbital evolution of a system of interest.
If a SN event occurs and the system is not disrupted the 
velocity injected into the system by a SN, calculated 
within \textsc{bse}, is simply 
vectorially added to the known Galactic velocity of the system 
centre of mass at the time of the SN.
If the system is disrupted the run away velocities of the two stars
-- as calculated in Section~\ref{s:binupdate} -- are, again, vectorially added
to their previous Galactic velocity (that of their system of origin).
In this way we are able to follow the complete Galactic orbital
history of a system (star or binary), even if it passes through 
two SNe and with (or without) binary disruption.
Because we assume no interaction between orbiting systems
we are able to evolve each system separately, one after another
(or in parallel).
A beneficial consequence resulting from this assumption is 
that \textsc{binkin} is faster to run than other dynamical 
codes, such as typical N-body codes (McGlynn 1984).
Of course, if one is dealing with compact stellar clusters dynamical
interactions between the stellar components are very important to follow.

\section{Pulsar population statistics}
\label{s:results}

We now describe the results of a series of population 
synthesis calculations that utilise our \textsc{binpop} 
and \textsc{binkin} modules to follow the stellar/binary 
and kinematic evolution of a population of binary stars 
to produce artificial Galactic pulsar populations.
The primary aim of this section is to predict scale heights and 
other kinematic characteristics for populations of pulsars, firstly assuming
that all pulsars can be detected.
We also compare the kinematics of our model pulsar 
populations with available kinematic tracers found 
from pulsar survey observations (such as those produced in 
Yusifov \& Kucuk 2004; Lorimer et al. 2006 and 
Hobbs et al. 2005).
For now we hold back from making direct statistically 
significant comparisons as this requires modelling of selection
effects which will be covered in detail in a companion paper
(the \textsc{binsfx} component as mentioned in 
Section~\ref{s:intro}).
Here we focus more on showing how 
modifying certain parameters affects the final pulsar 
population kinematics, in terms of scale heights and
space velocity distributions.

\begin{table}
 \centering
 \begin{minipage}{84mm}
  \caption{A summary of the \textsc{binkin} models used
within this paper and their main assumptions.
The first column provides a label for each model.
This is followed by the assumed Galactic radial distribution
of pulsar birth locations, $R_{\rm init}$, where Paczynski stands for the 
distribution suggested by Paczynski (1990; see our 
Equation~\ref{e:PacRdist}), Flat is the flat distribution
described in Section~\ref{s:method} and Yusifov \& Kucuk
is the birth distribution suggested in Yusifov \& Kucuk (2004; 
see our Equation~\ref{e:YKRdist}).
The third column gives the Galactic gravitational model used
(see Section~\ref{s:potentials}).
The next column is the maximum height with respect to the
Galactic plane, $|z_{\rm max}|$, that we consider for our pulsar distribution
and scale-height calculations, followed by our assumed value
of the SN kick dispersion, $V_\sigma$.
Note that Model G evolves a population of single stars
while all other models start with a population of binaries.
  \label{t:table1}}
  \begin{tabular}{crrrr}
  \hline
 Model  & \multicolumn{1}{c}{$R_{\rm init}$ distribution} & \multicolumn{1}{c}{$\Phi$} & \multicolumn{1}{c}{$|z_{\rm max}|$} & \multicolumn{1}{c}{$V_\sigma$} \\
\hline 
 A & Paczynski & P90 & $10~$kpc & 190 \\
 B & Flat & P90 & $10~$kpc & 190 \\
 C & Yusifov \& Kucuk & P90 & $10~$kpc & 190 \\
 C2 & Yusifov \& Kucuk & P90 & $2~$kpc & 190 \\
 C3 & Yusifov \& Kucuk & P90 & $20~$kpc & 190 \\
 D & Paczynski & KG89 & $10~$kpc & 190 \\
 E & Paczynski & Xue08 & $10~$kpc & 190 \\
 F & Paczynski & P90 & $10~$kpc & 550 \\
 G & Paczynski & P90 & $10~$kpc & 265 \\
\hline
\end{tabular}
\end{minipage}
\end{table}

The first step is to evolve a population of binaries within
\textsc{binpop}.
For this we proceed using our favoured model from Paper I 
(Model Fd).
This sets choices for \textsc{binpop} stellar and binary 
evolutionary parameters of: solar metallicity $Z = 0.02$; a
maximum possible NS mass of $3~M_\odot$ and $\alpha_{\rm CE} = 3$.
It also sets the following parameters governing pulsar evolution: 
$\tau_{\rm B} = 2000~$Myr; $k = 3000$; no propeller evolution; 
the initial pulsar period and magnetic field parameter selections 
linked to the strength of the SN velocity kick; the angular 
momentum accreted by the pulsar is variable; no electron capture SNe; 
and no beaming of pulsars (see Section~\ref{s:binpop} and Paper~I for details).
SN kicks using the \textsc{bse} prescription are given to NSs 
(see Section~\ref{s:binupdate}), 
while to keep the required number of 
models down to a minimum we only use the curvature radiation 
death line model (Harding, Muslimov \& Zhang 2002).
Unless otherwise stated we take $10^7$ binary systems with
initial parameters selected in the same manner as within Paper~I 
and with the limits of $5 - 80~M_\odot$ for primary mass,
$0.1 - 80~M_\odot$ for secondary mass and $1-30\,000~$days
for orbital period.
The Galactic age is assumed to be $10~$Gyr.
Each binary is evolved to this age and for those that create pulsars
the evolution history e.g. SN occurrence times and velocities, is saved as
input for \textsc{binkin}.

The next step is to take each of the \textsc{binpop} binaries
and follow their corresponding kinematic evolution in \textsc{binkin}.
For each binary a random birth age is assigned and the evolution
followed from this age up to the age of the Galaxy.
For this we begin by defining a standard model which we will call 
Model A.
This uses the Paczynski (1990) distribution for setting the initial
Galactic radial positions of the binaries (see Equation~\ref{e:PacRdist}) 
with a maximum initial height off the plane of $|z_{\rm maxi}| = 75~$pc. 
It also assumes the P90 form of the Galactic potential 
(see Equations~\ref{e:Pac1} and \ref{e:Pac2}) 
and sets $V_\sigma = 190~$km s$^{-1}$ (used within \textsc{binpop}) 
as the dispersion of the SN velocity kick distribution.
Further models arise due to variations of these choices and are listed
in Table~\ref{t:table1}.

For each model we examine the scale heights for a range of 
pulsar systems.
These are given in Table~\ref{t:table3}.
We take the scale height to be that distance in $|z|$ for 
which the number of stars within that distance is $63\%$ 
($\sim$ twice the e-folding distance) of the entire population.
The most prolifically observed pulsar system is what 
is known as a standard pulsar.
Here we define a standard pulsar as one which satisfies 
\begin{equation}
\label{e:standard}
\log{B} \geq -2.5\times \log{P} + 8.1.
\end{equation}
This equation artificially divides the `standard' pulsar `island'
from all other radio pulsars in the $B-P$ diagram (see Paper~I). 
We define a MSP to be a pulsar spinning more rapidly than
$P = 0.02~$s.
All other pulsars bridge these two pulsar types -- islands
within the $P\dot{P}$ plane (see also the description given 
in Paper~I).
We also distinguish between binary and isolated pulsars.
It is possible to compare our model results in a limited manner 
to observations.
To do this we make use of the ATNF Pulsar Catalogue which
provides us with $\sim 1610$ Galactic plane pulsars (we ignore
pulsars from the catalogue that have any association with 
another object, for example with a globular cluster or
external galaxy). 
Approximately  $1550$ of these are isolated.
Only $15$ standard pulsars are found in binary systems within
the Galactic disk.
Of the total observed pulsars there are $65$ MSPs of which $19$ are isolated.
We show the scale heights of the catalogue pulsars within 
Table~\ref{t:table2}.

\begin{figure*}
  \includegraphics[width=168mm]{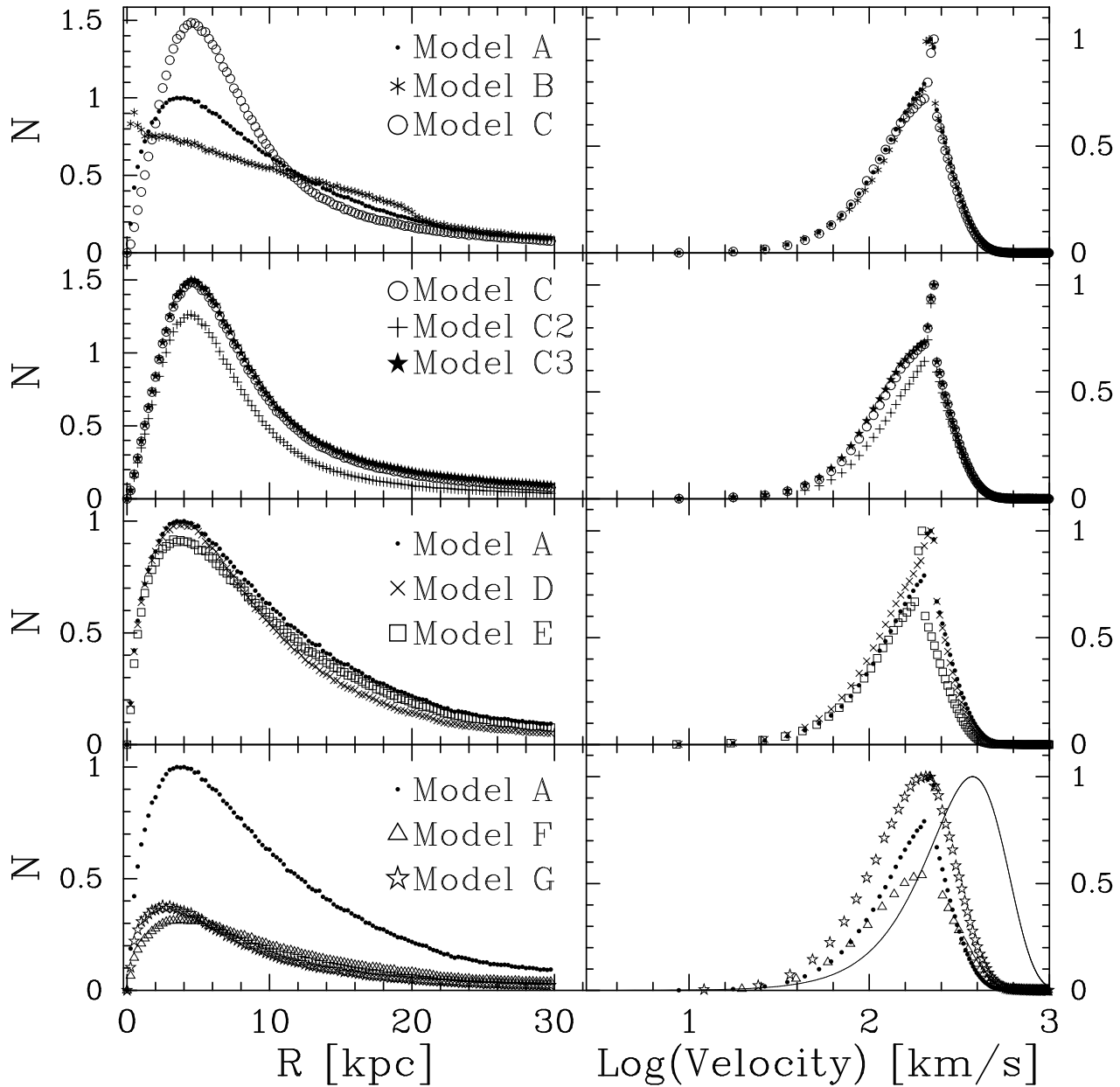}
  \caption{Left panels: theoretical radial distributions of pulsars 
for a range of models. 
All model radial distributions are normalised to Model A. 
Right panels: theoretical space velocity distributions for 
the same range of models. 
The velocity distributions depicted here represent the
pulsar population space velocities with respect to the 
Galactic centre.
The solid line in the bottom-right panel depicts the 
3D space velocity distribution of young pulsars derived from 
observations (Hobbs et al. 2005).
All velocity distributions are normalised to unity. 
   \label{f:fig8}}
\end{figure*}

The radial distributions of Galactic pulsars resulting from our set 
of models are shown in Figure~\ref{f:fig8} (left-hand side panels). 
We also compare a subset of the models in more detail in 
Figure~\ref{f:fig9} and include a comparison to the initial 
distributions used in the models and also the pulsar distribution 
suggested by Yusifov \& Kucuk (2004) which is based on observations. 
We also explore the pulsar population 3D space velocity distributions. 
These are shown in the right-hand side panels of Figure~\ref{f:fig8}
for the models in Table~\ref{t:table1}.

Our results are analysed in more detail in four parts.
In Section~\ref{s:initdist} we examine the effect of varying 
the assumed pulsar birth radial distribution.
This analysis makes use of Models A, B and C.
Within Section~\ref{s:initdist} we also consider how modifying
the target area considered (the `observable' Galactic area) 
in our scale height calculations affects the scale height values 
of pulsars produced in Model C.
This is completed with the use of Models C2 and C3.
Section~\ref{s:galpotmodels} analyses different forms of the 
Galactic gravitational potential by comparing Models A, D 
and E pulsar scale heights, final radial distributions and final 
velocity distributions. 
We then consider the effect of varying the assumed SN velocity kick 
distribution with Models A, F and G in 
Section~\ref{s:kickresults}.
Finally, after examining differences in bulk pulsar properties, we 
explore in detail the MSP population of Model C in Section~\ref{s:MSPsection}. 
We make use of our MSP analysis to further investigate the effect of 
model assumptions such as the initial scale 
height, Galactic age and the number of systems evolved. 

\begin{table*}
 \centering
 \begin{minipage}{140mm}
  \caption{Model scale heights for a range of pulsar types in the Galaxy.
  \label{t:table3}}
  \begin{tabular}{ccrrrrrrrrr}
  \hline
  & & Model & A & B & C & D & E & F & C2 & C3 \\
  \hline 
  \multicolumn{2}{c}{Type} & & & & & & & & & \\
  \hline 
  & Both & & $1.39$ & $1.49$ & $1.25$ & $1.33$ & $2.58$ & $1.93$ & $0.58$ & $1.40$ \\
  All & Isolated & & $1.43$ & $1.53$ & $1.30$ & $1.37$ & $2.64$ & $1.93$ & $0.59$ & $1.46$ \\
  & Binary & & $0.96$ & $1.10$ & $0.79$ & $0.91$ & $1.98$ & $2.00$ & $0.46$ & $0.83$ \\
  \hline 
  & Both & & $1.43$ & $1.53$ & $1.30$ & $1.36$ & $2.63$ & $1.93$ & $0.59$ & $1.45$ \\
  Standard & Isolated & & $1.43$ & $1.53$ & $1.30$ & $1.37$ & $2.64$ & $1.93$ & $0.59$ & $1.46$ \\
  & Binary & & $0.73$ & $0.85$ & $0.59$ & $0.71$ & $1.68$ & $1.58$ & $0.38$ & $0.61$ \\
  \hline 
  & Both & & $1.00$ & $1.15$ & $0.82$ & $0.95$ & $2.04$ & $2.02$ & $0.47$ & $0.88$ \\
  All MSPs & Isolated & & $1.76$ & $1.67$ & $1.76$ & $1.49$ & $2.90$ & $2.33$ & $0.60$ & $2.21$ \\
  & Binary & & $0.99$ & $1.14$ & $0.82$ & $0.94$ & $2.03$ & $2.02$ & $0.47$ & $0.87$ \\
  \hline   
\end{tabular}
\end{minipage}
\end{table*}

\subsection{Initial distributions and target area}
\label{s:initdist}

To begin with we focus on Model~A. 
We note that unless otherwise specified, when calculating scale heights 
we consider only pulsars located within the Galactic region defined by 
$|z| \leq |z_{\rm max}|~$kpc and 
$r = \sqrt{x^2 + y^2}  \leq 30~$kpc. 
Firstly comparing the scale heights of the isolated and binary pulsars 
we see that as expected (see Section~\ref{s:intro}) it is the 
isolated pulsars which have the greatest scale height when considering 
all pulsars.
It is also no surprise that the scale height of the complete population 
(top row of Table~\ref{t:table3}) is closely aligned with the isolated 
pulsar scale height. 
This is because 91\% of pulsars in Model A 
are isolated at the end of the simulation 
(even though all stars are initially in binaries). 
When considering only standard pulsars the domination of 
the isolated component is even greater: 
99\% of standard pulsars are isolated. 
However, the tables are turned when we look at the MSP population: 
the binary MSP population makes up 99\% of all MSPs within Model A
(a greater percentage than what is observed, however, see
Section~\ref{s:MSP1} for further discussion on this point). 
These relative numbers explain why in Table~\ref{t:table3} the 
scale height of standard pulsars is insensitive to 
standard binary pulsars, and the same can be said for the 
total MSP scale height compared to isolated MSPs. 

For the binary pulsars we find that the 
standard pulsar population has a 
lower scale height than for the MSP population. 
This suggests that on average binary MSPs receive greater 
post-SN recoil velocities than their standard binary pulsar 
counterparts. 
The reasons for this were alluded to in Section~\ref{s:intro} 
but we reiterate them here (see also the findings of 
Stollman \& van den Heuvel 1986; Bailes 1989). 
Basically, the very existence of an MSP relies on the occurrence 
of mass-transfer on to the NS which in turn requires a close binary. 
Such a binary will have a greater binding energy than the 
equivalent standard (wider) binary pulsar and can therefore survive 
a greater SN velocity kick as there is more energy to overcome 
for disruption. 
As a result it is possible for the proto-binary MSP system to 
be given a faster recoil velocity 
(with respect to the systems initial CofM). 
The isolated MSP population scale height also increases 
compared to the entire isolated pulsar population.
Again this is not surprising given the model isolated MSP 
formation scenario as addressed in Paper I and within 
Section~\ref{s:intro}. 
Owing to the SN that produced the NS progenitor of the MSP, 
and the SN of the companion star that disrupted the system, 
isolated MSPs can receive greater recoil velocities
than any other pulsar population -- hence the largest population
scale height (this formation mechanism
is also discussed further in Section~\ref{s:MSPsection}).

To measure what effect the initial Galactic birth distribution 
has on the final model pulsar population we compare Models A, B and C. 
These models assume the birth distribution of Paczynski (1990: from the 
observed distribution of SN remnants), a uniform thin disk, 
and the birth distribution of Yusifov \& Kucuk (2004: derived from 
observations of OB stars), respectively. 
The final radial distributions (at a Galactic age of $10\,$Gyr)
for the pulsars in these models are compared in the upper-left 
panel of Figure~\ref{f:fig8}. 
Both Models A and C have distributions that peak away from the 
Galactic centre (reflecting their initial distributions). 
However, the peak of Model A is shallower and the distribution 
is more extended than Model C. 
We note that the number of pulsar systems `observed' in Model C 
is higher than in Model A (by a factor of $\sim 1.1$). 
For Model B we find that the pulsar radial distribution peaks at the 
Galactic centre and then decays with increasing radius. 
The space velocity distributions for the pulsars in the three models 
are compared in the upper-right panel of Figure~\ref{f:fig8} and we 
see that there is no discernible difference. 

\begin{figure}
  \includegraphics[width=84mm]{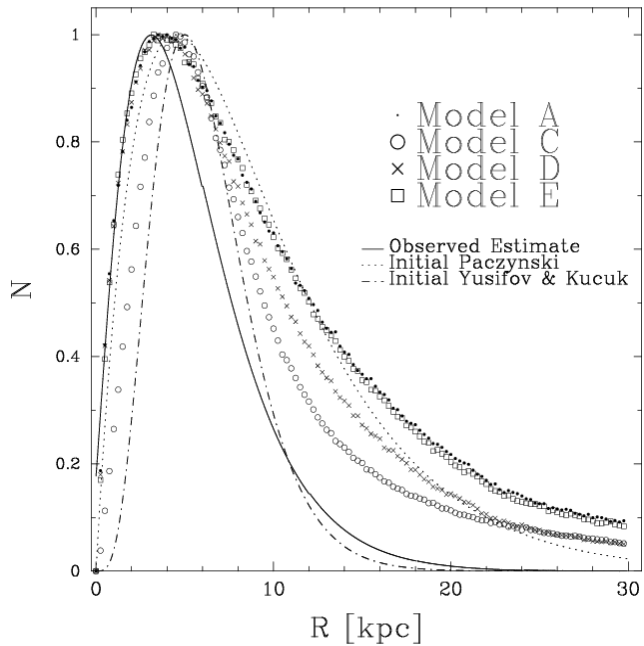}
  \caption{Theoretical radial distributions of pulsars 
for a select few models all normalised to unity. 
Here we restrict ourselves to only consider those pulsars with
$\dot{P} > 10^{-17}~$s s$^{-1}$ which compares to the observed sample
used by Yusifov \& Kucuk (2004: solid line). 
Also provided is the initial radial distribution given in 
Paczynski (1990: dotted line), which is used in Models A, D and E,
and the initial radial distribution of Yusifov \& Kucuk (2004: 
dash dotted line) used in Model C. 
   \label{f:fig9}}
\end{figure}

The relation of the final pulsar radial distribution to the assumed 
birth distribution of binaries can be seen in Figure~\ref{f:fig9} 
for Models A and C. 
Here all distributions are normalised to unity to aid comparison 
of the peak position and distribution shapes. 
We see that the birth distribution of Model A is broader than for Model C 
and this is reflected in their final shapes. 
However, the width of the distribution increases with time in both cases, 
while the peak of the distribution moves in towards the Galactic centre,
which is a typical effect of Galactic potentials (Sun \& Han 2004). 
Initially Model A peaks at a radius of $4.5\,$kpc which moves inwards 
to a radius of $3.9\,$kpc at $10\,$Gyr. 
For Model C the peak moves from $5.0\,$kpc to $4.5\,$kpc. 

In Figure~\ref{f:fig9} we also compare the model distributions 
to the distribution of an observed sample of pulsars presented 
by Yusifov \& Kucuk (2004). 
We note that at this stage we are not including selection effects 
in our models so a direct comparison with observations is not 
possible. 
However, comparison with the Yusifov \& Kucuk (2004) 
sample, which includes selection effects somewhat by being 
limited to pulsars with $\dot{P} > 10^{-17}~$s~s$^{-1}$, 
can still provide a meaningful guide to discerning between 
our models. 
Although not included in Figure~\ref{f:fig9} we can see 
immediately that Model B is an unrealistic model 
of the Galactic pulsar population. 
The apparent deficit of observed pulsars in the inner 
region of the Galaxy can not be reproduced by assuming 
all binaries are born in a uniform thin disk 
-- we require a paucity of pulsars 
to be born in the central region of the Galaxy 
when attempting to match observations (similar to
Paczynski 1990; Sun \& Han 2004).
This lack of observed inner Galactic pulsars may be due to
the high electron density in this region of the Galaxy and 
therefore larger scattering of the pulse signal, however,
the latest observations do suggest an intrinsic scarcity of central pulsars 
(Lorimer et al. 2006). 
Model A provides a good comparison to the observed radial pulsar distribution 
for the inner regions of the Galaxy but has too many pulsars and is too 
extended beyond $\sim 4~$kpc. 
In terms of shape, Model C best represents the observations. 
However, Model C peaks further from the Galactic centre by $1 - 1.5\,$kpc. 
This suggests that the ideal initial distribution would of the form derived by 
Yusifov \& Kucuk (2004) from observations of OB stars but scaled so that 
the distribution peaked at a radius of $\sim 4\,$kpc. 

The scale heights for Models A, B and C can be 
compared in Table~\ref{t:table3}. 
We see that Model B has systematically the largest 
scale heights while Model C has the smallest. 
However, the trends observed for Model A 
-- the relative scale heights of the various pulsar populations 
-- are consistent across  the models

We next demonstrate what effect modifying the Galactic 
region of interest has on Model C by considering pulsars
only out to $|z| = 2~$kpc in Model C2 and out to 
$|z| = 20~$kpc for Model C3 
(as opposed to $|z| = 10~$kpc for Model C). 
We find that reducing the height of our 
`Galaxy' by a factor of five approximately halves the 
calculated scale heights for all pulsar populations. 
On the other hand, doubling the height considered does not 
significantly affect the calculated scale heights (except 
perhaps the isolated MSP population which suffers from small number 
statistics). 
However, it does not appear to greatly affect the relative 
scale heights of pulsar sub-populations and certainly does 
not switch any trends noted in the models. 
Beyond this limit only the results of highly
energetic systems may still vary. 
Therefore, factors which limit the region of the 
Galaxy observed (or considered), such as the numerous
selection effects which occur in radio pulsar 
observations, can modify the underlying pulsar 
population scale heights within $|z| < 10~$kpc
of the Galactic plane (as discussed by many works including
Taylor \& Manchester 1977 and Narayan \& Ostriker 1990). 
We note that in terms of Galactic pulsar observations
there is the limit of $\sim 1.75~$kpc beyond which 
dispersion measure distance estimates of pulsars 
break down (see Manchester et al. 2005). 

We now have Model C, a suitable model for which we may compare
pulsar scale heights to observations (the latter values 
are given in Table~\ref{t:table2}).
Model C2 is roughly consistent with the observed scale heights, 
although on average the model values are greater
than the observed values. 
The trends when considering all pulsars are similar but this 
breaks down for the MSP population where the model predicts 
a greater scale height for isolated MSPs than for binary MSPs 
which is opposite to the observed MSP scale heights (although,
see Section~\ref{s:MSP1}). 
This is, however, consistent with our previous simple analysis in
Section~\ref{s:intro} from the binary disruption formation 
mechanism of isolated MSPs. 
Another difference between Model C2 and observations is the 
relative number of isolated to binary MSP systems.
In Model C2 $\sim 99\%$ of MSPs are found within a binary
system while a direct observational comparison shows 
$\sim 70\%$ of Galactic disk MSPs in binary systems.
These two differences between our model and observations 
indicate that our mechanism for producing isolated MSPs 
-- binary disruption in a SN event -- can not be the sole 
(or even dominant) production mechanism. 
We explore this line of thought further in 
Section~\ref{s:MSPsection}.

\begin{table}
 \centering
 \begin{minipage}{84mm}
  \caption{Observed scale heights (in kpc) for different types of pulsars.
    We have not taken account of selection effects when calculating
    these values.
    Standard pulsars are those pulsars which satisfy 
    Equation~\ref{e:standard}, while MSPs are those pulsars
    which have spin-periods $P \leq 0.02~$s.
    We note that there is uncertainty in some of these numbers owing to
    small number statistics and the clumpy distribution of the
    pulsars.
    For example, there are only $19$ isolated MSPs and the difference
    in height between the 12th and 13th most distant (in terms of
    height from the plane) is $0.07~$kpc while the average distance between 
    the first twelve is $0.02~$kpc. 
  \label{t:table2}}
  \begin{tabular}{crrr}
  \hline
 Type & All & Standard & MSP \\
  \hline
 Both & 0.40 & 0.39 & 0.41 \\
 Isolated & 0.41 & 0.39 & 0.26 \\
 Binary & 0.38 & 0.17 & 0.48 \\
  \hline 
\end{tabular}
\end{minipage}
\end{table}

\subsection{Model Galactic gravitational potentials}
\label{s:galpotmodels}

We now explore what effect modifying the assumed 
Galactic gravitational potential has on the pulsar 
scale heights and final distributions (radial and space velocity). 
Model D assumes a potential of the form described 
by the KG89 model (Equation~\ref{e:KGpot}).
The scale heights of Model D are all slightly less 
than their Model A counterparts which used the P90 model. 
This suggests that over time stellar systems may diffuse less 
efficiently in Model D than in Model A. 
Figure~\ref{f:fig4} gives some indication of the cause of this difference. 
The lower panel of provides the model Galactic gravitational 
force towards the plane with respect to the height above the plane. 
We see that above a height of $|z| \sim 1~$kpc the KG89 model 
has a slightly greater attractive force than the P90 model. 
However, for the inner regions the situation is reversed and 
indeed if we calculate the scale heights for pulsars with 
$|z| < 1~$kpc, we find that the scale height behaviour for 
Model D relative to Model A is also reversed. 
As shown in Figures~\ref{f:fig8} and \ref{f:fig9} the 
radial distribution does not differ greatly between Models 
A and D:  small differences to note are that the distribution of 
Model D decays more rapidly 
than Model A while Model D contains less pulsar systems 
within $10~$kpc of the Galactic plane.
The velocity curve of Model D does not change greatly from 
Model A as there is only a small increase in systems with 
space velocities at the lower end of the distribution, perhaps 
reflecting the rapid decay in circular velocity of Model D
with respect to Model A (see Figure~\ref{f:fig4}).

Model E assumes the Galactic gravitational potential of 
Xue08 (Equations~\ref{e:Xue1} to \ref{e:Xue3}).
This model shows a large increase in scale height values compared to 
both Models A and D.
The reason for this can once again be seen from 
Figure~\ref{f:fig4} which shows that 
the Xue08 Galactic gravitational model
can not retard the movement of systems out of the Galactic 
plane as effectively as the P90 (Model A) or KG89 (Model D) models. 
Figure~\ref{f:fig8} shows that the number of systems retained
by Model E within $|z| \leq 10~$kpc is less than within
both Models A and D.
The 3D space velocity curves of Models A, D and E all peak 
at roughly the same value but the Model E distribution possesses 
a much steeper slope either side of its peak value. 
The flatness of the rotation curve assumed in Model E, depicted
in Figure~\ref{f:fig4}, causes such a narrow distribution in 
velocity space.
Model E highlights the importance of the assumed Galactic 
gravitational potential in modelling Galactic population 
kinematics (as also discussed by Kuijken \& Gilmore 1989, 
Dehnen \& Binney 1998 and Sun \& Han 2004).

\subsection{Kicks}
\label{s:kickresults}

Considering the uncertainty involved in the true form of the 
distribution of kick speeds given to NSs at birth (as mentioned 
in Section 2.2.1) 
we next investigate how changing the dispersion of the assumed 
Maxwellian distribution affects our results. 
Recalling that Model A used $V_\sigma = 190~$km s$^{-1}$ we first 
compare with Model F which uses $V_\sigma = 550~$km s$^{-1}$ 
as an extreme illustration. 
We see from Table~\ref{t:table3} that the scale heights in Model F 
are much greater than in Model A, with increases by as much as 
a factor of 2. 
While it is expected that pulsars can move further from
the galactic plane in Model F it also means that more 
objects escape from the Galaxy: 
the number of pulsars in Model F within $10\,$kpc of the plane decreases 
by a factor of 10 compared to Model A. 
This decrease in numbers can skew the expected outcomes and 
is evident when looking at the radial distributions 
in Figure~\ref{f:fig8}. 
For example, less binary systems are kicked out of the 
$|z| \leq 10~$kpc Galactic region than isolated pulsars 
-- owing to binaries being heavier on average and the binary orbit 
absorbing a fraction of the energy injected by the kick 
-- so we find in Model F that binary pulsars have a greater 
scale height than isolated pulsars (the opposite to Model A). 
This is not the case for MSPs although the difference between 
the isolated and binary MSP scale heights has decreased compared 
to Model A. 
We note that the ratio of standard isolated pulsars to 
standard binary pulsars is an order of magnitude greater 
for Model F than for Model A owing to a greater incidence 
of binary disruptions in Model F. 
Therefore, as established in previous works 
(e.g. Taylor \& Manchester 1977; Hills 1983), the assumed
SN kick velocity is an extremely important factor, especially
when comparing model pulsar kinematics to observations.

When examining the resultant radial distribution in Figure~\ref{f:fig8} 
we find that Model F is much broader than Model A 
-- an intuitive results owing to the increased distance 
that the pulsars move -- and certainly Model F is not a 
good representation of the observed 
distribution (not that it was expected to be). 
Comparing the space velocity distributions of Models A and F 
in Figure~\ref{f:fig8} it is surprising to see the relatively high 
number of pulsars in Model F that are traveling at the distribution 
peak speed ($\sim 250~$km s$^{-1}$). 
This population with similar velocities includes isolated and 
binary pulsars. 
We remind the reader that when discussing the velocity
distribution here we mean the actual velocity each
system has with respect to the Galactic centre --
we do not account for the local standard of rest 
(solar motion). 

In Figure~\ref{f:fig8} we also compare the 3D space velocity distributions 
of Models A and F with the 3D space velocity distribution 
derived by Hobbs et al. (2005) from pulsar observations. 
An important distinction to make is that the Hobbs et al. (2005) sample was 
restricted to pulsars with characteristic ages $< 3~$Myr.
Thus, it is intended to be a distribution of pulsar birth
velocities and was used by Hobbs et al. (2005) to suggest
that $V_\sigma = 265~$km s$^{-1}$ in the SN velocity kick distribution. 
By comparing this with our models we can gauge the effect that the Galactic 
potential and binarity have on the form of the pulsar velocity distribution 
as the population evolves. 

Comparing the Model A pulsar velocity distribution (at a population age 
of $10\,$Gyr) with the Hobbs et al. (2005) birth velocity distribution 
shows changes in the peak (shifted to lower velocity in the model) and 
shape. 
The shift of the peak can be mostly attributed to the difference in the 
average age of the two pulsar populations 
and the low dispersion value assumed in the Model A velocity 
kick distribution. 
The age difference allows fast moving systems in Model A to have 
time to leave the Galaxy and thus be culled 
from the final velocity distribution.
Also, over time, the pulsar velocities are retarded by the Galactic
potential which shifts the final pulsar velocity distribution to 
lower velocities. 
In terms of shape we find that Model A is a more focussed 
distribution -- the model peak is more acute and the distribution as a 
whole is narrower.
This difference is likely due, in part, to the binding energy 
of the host binaries impinging on the SN velocity kick 
(as discussed in Hills 1983 and Bailes 1989) and causing a greater 
number of systems to have similar final space velocities than otherwise.
Disruptions triggered primarily by mass loss will act to increase the 
proportion of low velocity pulsars while the absorption 
of large kick velocities by the binary binding energy may also skew 
the distribution to smaller final pulsar velocities. 
This narrowing of the model pulsar velocity distribution compared
to the observed distribution has been found in other works,
most notably that of Dewey \& Cordes (1987).
They attributed the difference to errors in pulsar distance
measurements, which will broaden the distribution, and that
non-Maxwellian processes may be more important in producing 
pulsar velocities than their models assume (that is, nascent 
NS receiving a kick selected from a Maxwellian distribution 
with $<V> = 90~$km s$^{-1}$).
It appears that the difference between the velocity distribution of 
models and observations results from a combination of the selected 
pulsar population used to derive the observed pulsar velocity 
distribution (see Hobbs et al. 2005), errors in pulsar velocity and 
distance measurements, and the binding energy of host binary systems
affecting the resultant pulsar run-away velocity.

To remove the binary orbit effect and highlight the effect of age 
evolution on the pulsar velocity distribution 
we have created Model G which evolves a population of single 
stars according to the Galactic setup described for Model A but with 
$V_\sigma = 265~$km s$^{-1}$. 
With every system evolved within Model G being isolated from birth 
we are now able to compare the resultant velocity distribution of a
population of pulsars which receive uninhibited SN kick velocities
drawn directly from the suggested Hobbs et al. (2005) SN kick distribution.
We now see in Figure~\ref{f:fig8} that the distribution closely resembles 
the Hobbs et al. (2005) distribution in shape but is shifted to 
lower velocities. 
The final pulsar distribution is best fit by a Maxwellian distribution 
with $V_\sigma = 140~$km s$^{-1}$. 

\subsection{Millisecond pulsars}
\label{s:MSPsection}

\begin{table}
 \centering
 \begin{minipage}{84mm}
  \caption{Scale heights (in kpc) of MSPs and 
    their companions for Model C (see Section 4.4.1) 
    and its variants (see Sections 4.4.4 and 4.4.5). 
  \label{t:table4}}
  \begin{tabular}{crrrrrrr}
  \hline
 Model & C & C$^{'}$ & C$^{''}$ \\
  \hline
 MSP-MS & $0.74$ & $0.75$ & $0.74$ \\
 MSP-WD & $0.83$ & $0.84$ & $0.86$ \\
 MSP-NS & $1.96$ & $1.93$ & $2.06$ \\
 MSP-BH & $0.11$ & $0.13$ & $0.10$ \\
 Isolated MSP & $1.76$ & $1.76$ & $1.61$ \\
  \hline
 Ablation & C & C$^{'}$ & C$^{''}$ \\
  \hline
 Binary MSP & $0.87$ & $0.87$ & $0.91$ \\
 Isolated MSP & $0.75$ & $0.75$ & $0.77$ \\
  \hline
  \end{tabular}
  \end{minipage}
\end{table}

In Paper I we were primarily interested in the production 
of MSPs within the $P$-$\dot{P}$ diagram. 
We now continue our exploration of the MSP population 
by examining in more detail the Galactic MSP distributions 
and in particular focussing on the behaviour of isolated MSPs 
and those with MS star, WD, NS or BH companions. 
In doing this we focus solely on Model C. 
To begin, we extend our evaluation of scale heights in
Table~\ref{t:table3} with those of the MSP populations 
(given in Table~\ref{t:table4} and discussed in Section~\ref{s:MSP1}).
The scale heights in Table~\ref{t:table4} are supplemented
by Figure~\ref{f:fig10} which provides the scale height for 
each MSP population as a function of Galactocentric radius.
Also shown is the Galactic $x$ and $z$ parameter space of MSPs: 
for all MSPs (see Figure~\ref{f:fig11}) and those that reside above 
a magnetic field cut-off (see Figure~\ref{f:fig12}). 
The population of MSP-BH binaries is then discussed in detail
within Section~\ref{s:MSP2}.
We then look at the MSP population recoil velocities and space velocities
in Section~\ref{s:MSP3} and make some limited comparisons 
to previous work and observations. 
To further our investigation into how different
model assumptions affect our pulsar population kinematics, 
we modify Model C, our favoured model thus far, 
to account for: a greater birth $|z_{maxi}|$ range (Model C$^{'}$
in Section~\ref{s:assinitsh});
a greater age of the Galaxy (Model C$^{''}$ in 
Section~\ref{s:galage}); and a higher resolution 
sample 
(in Section~\ref{s:resolution}).

\subsubsection{Model C MSP scale heights and Galactic spatial properties}
\label{s:MSP1}

Looking at the scale height values in Table~\ref{t:table4}
for Model C we see that as expected the binary MSP population 
with the greatest scale height is the MSP-NS
systems, in which two SNe kicks occur.
These double compact systems, however, are much rarer than 
the MSP-MS or MSP-WD systems and therefore the results
are less statistically significant.
The relative numbers of MSP-NSs compared to 
MSP-WD systems (the largest MSP population) is $0.003$.
For MSP-MS systems (the second most numerous MSP population) 
the relative number is $0.044~$per MSP-WD system.
We find similar scale heights for the MSP-MS and MSP-WD systems 
although the former are systematically smaller owing to the population 
being younger on average.  
Recently there have been suggestions that asymmetric mass-loss during 
the asymptotic giant branch phase (Spruit 1998) give rise to WD recoil 
velocities of the order of a few ${\rm km} \, {\rm s}^{-1}$ 
(Fellhauer et al. 2003). 
Such kick velocities have been raised a possible explanations of the 
apparent deficit of WDs in open clusters (Fellhauer et al. 2003) 
and the radial distributions of WDs in globular clusters 
(Heyl 2007; Davis et al. 2008).  
Currently we do not include this possibility in our models but note that 
it would presumably lead to a modest increase in the MSP-WD population 
scale height. 

MSP-BH systems are found to have a small Galactic
scale height.
This arises due to the orbital parameters required 
in order to form these systems which we examine in 
further detail below (see Section~\ref{s:MSP2}).
Also, we remind the reader that we currently assume 
BHs do not receive kicks during their formation.
As shown in Tables~\ref{t:table3} and \ref{t:table4} 
the isolated MSP population has a scale height of $1.76~$kpc.
These MSPs emerge from disrupted binary systems and although
the kick at the time of disruption may be large
it is not the MSP which is exploding at that point. 
Therefore the MSP is considered by our kick 
routine to be the secondary star which, as shown 
in Section~\ref{s:modelcomp}, receives (on average)
only a small increase in momentum. 
This results in the lower scale height 
of isolated MSPs compared to MSP-NS binary systems (albeit
only slightly less than the MSP-NS value).
Furthermore, we note that for the binary system to 
survive the first SNe, allowing mass transfer onto 
the progenitor MSP, the resultant velocity kick at 
this point must be relatively small (we find $V_{\rm kick}$ 
of approximately $80~$km s$^{-1}$ or less).
This is in accordance with many other population synthesis
works, including Stollman \& van den Heuvel (1986),
Iben, Tutukov \& Yungleson (1995) and 
Ramachandran \& Bhattacharya (1997).

Previous results shown in Section~\ref{s:initdist}
placed doubt on isolated MSPs formed via the disruption 
of binary systems being the sole `type'
of isolated MSPs -- there must be another formation 
mechanism. 
One such mechanism that exists in the literature 
is the ablation model (Eichler \& Amir 1988; 
Ruderman et al. 1989) based on observations such as
those of van Paradijs et al. (1988). 
Here the assumption is that the MSP is produced as a result of mass-transfer 
from a MS companion in what would be a low-mass X-ray binary. 
Then at some point the mass of the MS star becomes low enough that 
it is destroyed, or ablated, by the highly energetic radiation flowing  
from the rapidly spinning pulsar (van Paradijs et al. 1988;
Tavani 1992). 
We calculate that the timescale for the destruction of the MS 
companion star in this manner 
should take of order $\sim 5~\rm{Myrs}$ once the 
companion is below a mass of $\sim 0.02\, M_\odot$ 
(see Appendix~\ref{s:isomsp}).
Thus we propose a simple model to belatedly estimate the impact 
of ablation on our results where we assume that any MSP with 
a MS companion of mass less than $0.008 M_\odot$ (to be on the safe side) 
is in fact an isolated MSP. 
With the inclusion of ablation we find that the percentage 
of isolated MSPs increases from 1\% to 36\%. 
This new value is in rough agreement with observations 
where it is estimated that one third of the MSPs are isolated 
(Huang \& Becker 2007). 
Iben, Tutukov \& Tungleson (1995) similarly found 
good agreement with observations for binary to isolated 
ratios when assuming ablation of MSP companions.
We see from the last two rows in Table~\ref{t:table4} that 
the isolated and binary MSP populations now have comparable 
scale height values (in fact the isolated scale height is now 
slightly the lower of the two). 
Therefore the kinematics of the binary and isolated MSP 
populations are now similar.
This last point is actually consistent with 
the observations of MSPs, which via statistical
arguments show no difference in binary and isolated 
MSP kinematics (Lorimer et al. 2007).
From this simple test we see that the 
low mass companions to MSPs do occur and that the 
ablation process deserves serious consideration in 
future models.

\begin{figure}
  \includegraphics[width=84mm]{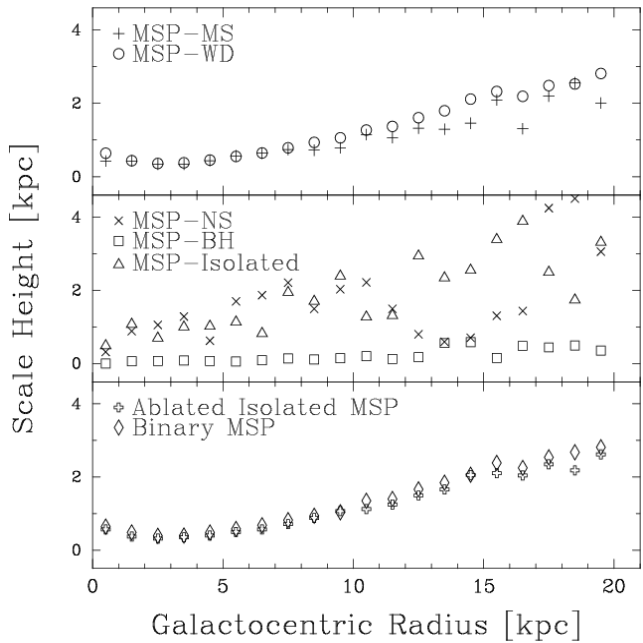}
  \caption{The radial variation of pulsar scale heights 
    for a range of MSP populations in Model C. 
   \label{f:fig10}}
\end{figure}

In Figure~\ref{f:fig10} we show how the scale heights of the MSP 
populations vary with Galactocentric radius. 
We note that the region of the Galaxy where the populations are 
most numerous is between $4-6~$kpc from the Galactic centre.
The top panel of Figure~\ref{f:fig10} depicts 
the similarity of MSP-MS and MSP-WD kinematics.
It is only out beyond $\sim 13~$kpc that the two
populations diverge, and this is only due to
low number statistics which begin to plague the MSP-MS
results. 
Low number statistics have a much greater influence in
the middle panel of Figure~\ref{f:fig10}.
For example, the highest number of systems 
in a radial bin ($1~$kpc in width) for 
the MSP-NS population contains $39$ systems
while the lowest only $3$.
The MSP-NS and isolated MSP systems have similar
scale heights throughout the majority of the 
Galaxy (after accounting for statistical uncertainty) and 
systems can be found far from the plane. 
The MSP-BH systems on the other hand are all 
found close to the Galactic plane.
When accounting for the ablation of MSP companions
we find a very similar distribution
of binary and isolated MSPs throughout the entire
Galaxy. 

\begin{figure}
  \includegraphics[width=84mm]{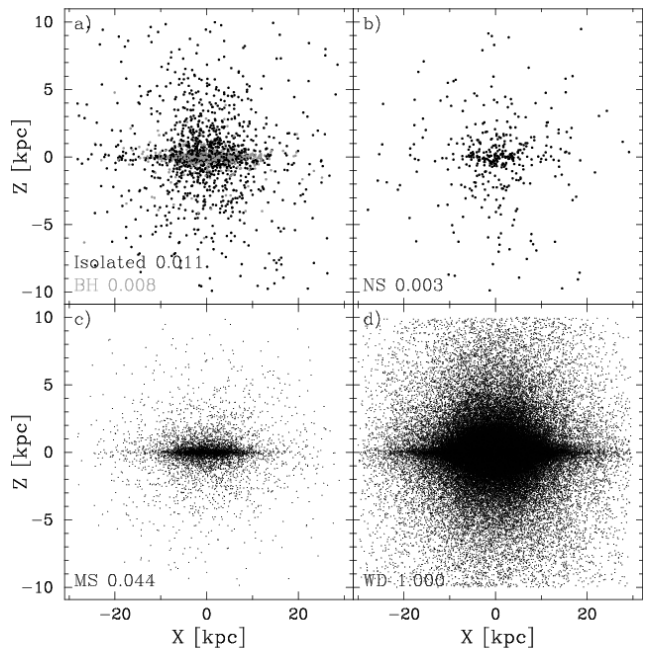}
  \caption{Galactic $x$ and $z$ coordinates for 
    Model C MSP systems within a radius 
    $R = \sqrt{x^2 + y^2} < 30~$kpc of the 
    Galactic centre and $|z| \leq 10~$kpc from 
    the Galactic plane.
    The bottom left corner of each panel gives
    the MSP companion type and relative number
    of that system compared to the MSP-WD systems.
    Due to their lack of numbers the points for
    MSP-NSs, MSP-BHs and isolated MSPs are larger
    than for the MSP-MS and MSP-WD systems.
   \label{f:fig11}}
\end{figure}

\begin{figure}
  \includegraphics[width=84mm]{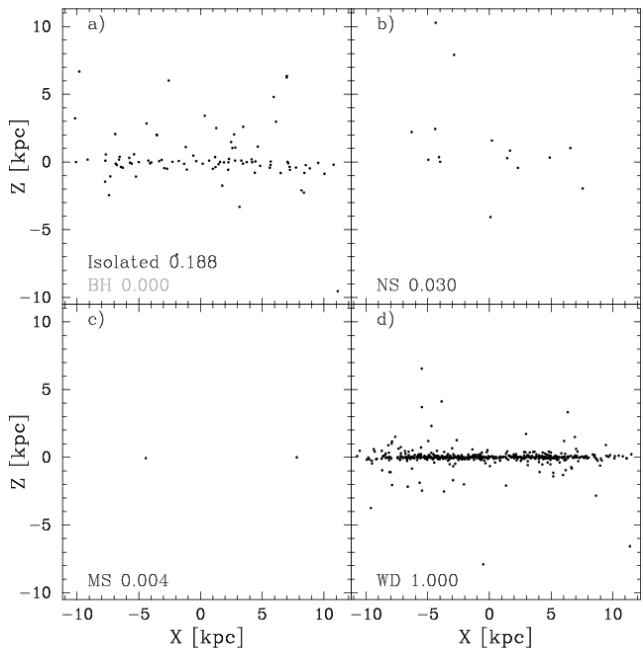}
  \caption{As for Figure~\ref{f:fig11} but now restricted 
  to include only pulsars that satisfy $B \geq 6\times 10^7~$G. 
   \label{f:fig12}}
\end{figure}

It is also interesting to compare the spatial 
Galactic $x-z$ distributions of the MSP populations.
This is shown in Figure~\ref{f:fig11} for the Galactic $xz$-plane 
and emphasises what we have already seen in Table~\ref{t:table4}
and Figure~\ref{f:fig10}: isolated
MSPs and MSP-NS binaries have quite extended 
distributions (relative to their numbers), 
MSP-BH systems reside close to the plane 
and the majority of MSPs are found with 
WD companions (MSP population numbers relative 
to MSP-WD systems are given in the 
lower right corner of each panel). 
What is surprising in Figure~\ref{f:fig11}
is the large number of MSP-WD systems out to 
$|z| = 10~$kpc.
This suggests that there may be many MSP-WD
systems lost from -- but surrounding -- the 
Galaxy. 
We next investigate the result of imposing a limited selection 
effect on the MSP population where we only consider pulsars 
that have $B > 6\times10^7~$G. 
This magnetic field value is a suggested 
limit (Zhang \& Kojima 2006) of the required field strength 
to turn on (or off) the pulse mechanism 
(see Paper I). 
Figure~\ref{f:fig12} shows the field strength limited MSP 
population and the result in comparison to Figure~\ref{f:fig11} 
is dramatic. 
The entire MSP-BH population now disappears, 
which is also almost the case for the MSP-MS 
population where only two systems are left.
The relative numbers of both isolated MSPs 
and MSP-NSs have now increased compared to 
the MSP-WD systems (see values on 
Figures~\ref{f:fig11} and \ref{f:fig12}).
Clearly many MSPs in Model C accrete enough mass 
to cause a large decay in the magnetic field.
In particular every pulsar within a MSP-BH system 
has accreted more than $\sim 0.04~M_\odot$ of material 
which is the typical amount of mass it takes for our model 
pulsar magnetic fields to decay below $B = 6\times10^7~$G (see Paper I).
This is compared to other works which assume $\Delta M > 0.1~M_\odot$ 
is required for MSP production (e.g. Willems \& Kolb 2005).

\subsubsection{Model C MSPs and BHs}
\label{s:MSP2}

\begin{figure}
  \includegraphics[width=84mm]{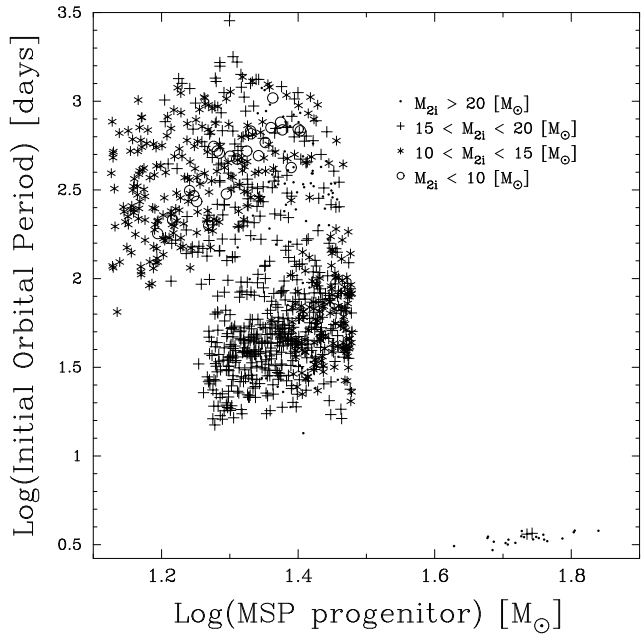}
  \caption{MSP-BH initial population parameter space of Model C,
    in particular the initial orbital period and 
    zero-age main-sequence primary star (MSP progenitor)
    mass.
    Provided are ranges of the initial companion mass
    (BH progenitor).
    Those systems with initial orbital periods, 
    $P_{\rm orbi} > 100~$days end with final orbital periods
    $P_{\rm orb} > 10000~$days.
    Those with $P_{\rm orbi} < 10~$days end with
    $20 < P_{\rm orb} < 10000~$days and those with
    $10 < P_{\rm orbi} < 100~$days end with
    $P_{orb} < 20~$days.
   \label{f:fig13}}
\end{figure}

Although other works have detailed BH and pulsar binary
evolution in varying detail (Narayan, Piran \& Shemi 1991;
Lipunov et al. 1994; 
Pfahl, Podsiadlowski \& Rappaport 2005; 
Lipunov, Bogomazov \& Abubekerov 2005) 
we evaluate the accretion history of MSP-BHs and why these systems 
reside close to the Galactic plane.
To place this into context we explore the initial 
orbital period and initial primary mass 
(the MSP progenitor and initially the more massive star) 
parameter space in Figure~\ref{f:fig13}.
This figure is also designed to show which systems 
reside in a range of \textit{observable} orbital periods 
(that is orbital periods which would be observed now, 
if the age of the Galaxy is $10~$Gyr).
Figure~\ref{f:fig13} also gives the secondary mass 
(BH progenitor and initially the less massive star) range
for each binary system depicted.
What we find in Figure~\ref{f:fig13} is three distinct 
groups in both initial orbital period and final orbital 
period.
Each of the three regions have different evolutionary pathways 
which leave the systems close to the Galactic plane.
We now describe these evolutionary pathways.
Firstly, however, we note the healthy number of MSP-BH systems
(few hundred) produced within our model.
This is in contrast to Pfahl, Podsiadlowski \& Rappaport (2005)
who suggest that BHs with recycled pulsar companions are rare.
The primary differences between our model and that of 
Pfahl et al. (2005) -- in terms of BH formation -- is their 
assumption that SN kicks are given to BHs, the assumed 
evolutionary parameter values within the CE phase and 
massive star wind mass-loss and accretion.
In varying the CE parameters Pfahl et al. (2005) show that
$\sim 100$ MSP-BH systems can be produced (with extreme CE 
parameter choices), however, a favoured model estimates only $5-10$ 
MSP-BH systems within the Galaxy.
This compares well with the number of systems we find with
orbital periods less than $\sim 10\,$hr, which is roughly the maximum
orbital period limit Pfahl et al. (2005) find for MSP-BH systems.
We also include an evolutionary pathway which forms an MSP-BH 
system that does not include a CE phase, something not considered
by Pfahl et al. (2005; who do, however, comment on this scenario).

Those systems which begin their lives with initial orbital periods 
less than $10\,$days all have an initial primary mass of 
$M_{\rm 1i} > 40~M_\odot$ and an initial secondary mass of 
$M_{\rm 2i} > 20~M_\odot$.
These systems end with the largest BH masses (around 
$\sim \,13~M_\odot$) of all the MSP-BH populations and their 
orbital periods are grouped near $1000~$days.
These systems were not found by Pfahl et al. (2005) possibly
owing to the inclusion of SN kicks during BH formation.
The general evolution pathway of these systems goes as follows. 
The initial orbital separation is of the order of $80\, R_\odot$ 
or less and the massive primary star evolves to fill its Roche-lobe 
within a few Myr. 
This leads to a phase of steady mass transfer lasting $1-2\,$Myr 
and ending with the primary as a naked helium star with a mass 
of about $10\, M_\odot$. 
During the phase the secondary accretes approximately 80\% of 
the transferred material with the remainder lost from the system. 
The orbital separation at this point is typically $200~R_\odot$ 
and subsequently increases further owing to winds from the helium 
star and the now massive secondary. 
At a system time of $\sim 5\,$Myr the primary undergoes a SN 
explosion and becomes a NS. 
We find that velocity kick magnitudes of $V_{\rm kick} \leq 80\,$km s$^{-1}$ 
allow the system to remain bound. 
Beyond the first SN the secondary evolves quickly and loses a large proportion 
of its matter in a wind, some of which is accreted by the NS companion. 
The secondary evolves via a naked helium star phase to explode as 
a SN and leave a BH remnant. 
At this point we have an eccentric MSP-BH system which has 
received one mild SN velocity kick in its lifetime and has typical 
component masses of 2 and $10-13\, M_\odot$, for the NS and BH respectively. 
The orbital separation is in the range of $1000 - 4000\, R_\odot$ 
(depending on the precise details of the kick velocity and the 
mass-loss history). 

Those MSP-BH systems with $10 < P_{\rm orbi} < 100\,$days as
seen in Figure~\ref{f:fig13} end their lives
with a large range of BH masses extending from $3~M_\odot$ 
through to $11~M_\odot$ in tight orbits around their MSP companion 
($P_{\rm orbf} < 20~$days).
It is this population of MSP-BHs which are most likely to coalesce 
at and around the age of the Galaxy (similar to Pfahl et al. 2005).
Initial primary masses are in the $18 < M_{\rm 1i}/M_\odot < 30$ 
range and secondary masses are typically $10 < M_{\rm 2i}/M_\odot < 20$. 
The initial orbital separation ranges from $100 - 300 R_\odot$. 
Early evolution proceeds similarly to that of the previous group: 
non-conservative mass transfer from the primary to the secondary 
accompanied by an increase in the orbital separation and ending with 
the primary as a naked helium star. 
The primary then undergoes a SN and becomes a NS at a system time 
of about $8\,$Myr. 
We find that generally these systems can survive slightly larger 
SN velocity kicks than the systems described in the previous group. 
The companion is now a massive MS star ($\sim 30\, M_\odot$) and 
subsequently fills its Roche-lobe while crossing the Hertzsprung Gap. 
This initiates dynamical-timescale mass transfer leading to a 
common-envelope phase and the creation of a tight binary comprised 
of the NS primary ($\sim 2\, M_\odot$) and a naked helium star 
secondary ($\sim 10\, M_\odot$). 
We note that systems in the first group avoid this second Roche-lobe 
filling event because the secondary is more massive and loses 
mass in a wind at a greater rate leading to more substantial orbit 
expansion after NS formation. 
After emerging from the common-envelope the NS then accretes material 
from the wind of the companion to become a MSP. 
This ends when the companion becomes a BH. 
The final MSP-BH binary will have an orbital separation of less 
than $10\, R_\odot$ and systems such as this may coalesce within 
a Hubble time. 

\begin{figure}
  \includegraphics[width=84mm]{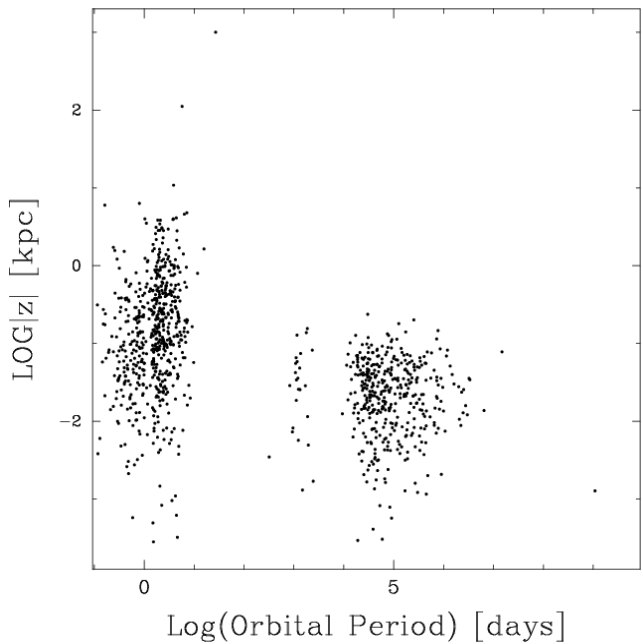}
  \caption{Model C MSP-BH pulsar $|z|-P_{\rm orbf}$ 
    parameter space.
    There are three distinct groups in final orbital period.
    The group on the left corresponds to systems with
    $10 < P_{\rm orbi} < 100~$days, the central group
    corresponds to systems with $P_{\rm orbi} < 10~$days
    and the right hand group corresponds to systems
    with $P_{\rm orbi} > 100~$days (see Figure~\ref{f:fig13}).
   \label{f:fig14}}
\end{figure}

The MSP-BH systems with $P_{\rm orbi} > 100\,$days end with orbital 
periods of 1000 days or more (see Figure~\ref{f:fig14} for the 
final orbital period range). 
We note that the smallest primary and secondary masses belong to 
this group. 
Once again mass-transfer occurs prior to the first SN event but 
owing to the wider orbit this is initiated later ($\sim 15\,$Myr) 
than in the previous cases and when the primary is a giant star. 
The orbital separation when the primary undergoes a SN (to become 
a NS) is typically $2000 - 3000\, R_\odot$ which means that relatively 
smaller kicks are required if the system is to remain bound and 
proceed to become a MSP-BH binary. 
We find that kicks of the order of 20 km s$^{-1}$ or less are necessary 
(slightly larger if the kick is well directed). 
After NS formation the secondary is a MS star with mass of 
approximately $20\, M_\odot$. 
The secondary then evolves off the MS and transfers some material 
to the NS before ending its life as a BH of mass less than $5\, M_\odot$. 

The above analysis shows that the most likely MSP-BH systems 
to be created are those in which the first SN -- the only 
one assumed to impart a velocity kick onto the compact remnant -- 
produces a small velocity kick, which is why 
these systems are found to hug the Galactic plane as suggested
by Narayan et al. (1991).
In fact, compared to the other MSP binary populations the MSP-BH systems 
effectively represent a different kick distribution, in that
the distribution of kicks given to systems that remain bound
is distinct.
As touched on in the evolutionary descriptions this is also 
true internal to the MSP-BH population, where
the effective kick distribution for systems that remain bound
is different for each of the three period groupings we 
identified in Figure~\ref{f:fig13}.
This is depicted indirectly in Figure~\ref{f:fig14}.
Here we see the MSP-BH height from the Galactic plane and 
the populations are designated 
by their grouping in the final orbital period parameter space.
Each population has a different scatter in $|z|$, which can be used 
as an indicator for the average strength of the SN velocity kick.
We see that the majority of those small orbital period MSP-BH 
systems are further off the plane than the extremely long 
period MSP-BH systems, suggesting that as expected from Bailes (1989),
the close binary systems can survive larger kick velocities than the larger 
binary systems (which was outlined in the evolutionary examples).
Only three MSP-BH systems are found beyond 
$10~$kpc from the Galactic plane.

The MSP-BH orbital period distributions as shown in 
Figure~\ref{f:fig14} are remarkably distinct and perhaps
surprisingly not smeared out by our use of random birth ages.
This is due to the vast orbital period differences between 
these populations and the time scales these populations evolve on.
The orbit of MSP-BH binary systems, after the formation of the BH, can only 
shrink in time owing to gravitational radiation 
(Landau \& Lifshitz 1951; Hulse \& Taylor 1985; Hurley, Tout \& Pols 2002). 
However, the time-scale on which this decrease occurs is 
greatly dependent on the size and eccentricity of the orbit.
Long period binary systems have very large timescales for orbital 
parameter change and thus remain as long period systems over 
a Hubble time. 
The very close systems (separation $\leq 10~R_\odot$) 
will shrink more rapidly and may even coalesce within a Hubble time. 
Therefore, the long period systems stay long and the short period 
systems only get shorter and as a result the MSP-BH systems stay 
within their orbital period groups as they evolve throughout the Galaxy. 
Thus we observe three distinct MSP-BH groups, a result differing 
some what from the orbital period distribution of 
Pfahl et al. (2005) who find that most MSP-BHs have orbital 
periods of $1-6$ hr.

\subsubsection{Model C MSP recoil and 3D space velocities}
\label{s:MSP3}

\begin{figure*}
  \includegraphics[width=168mm]{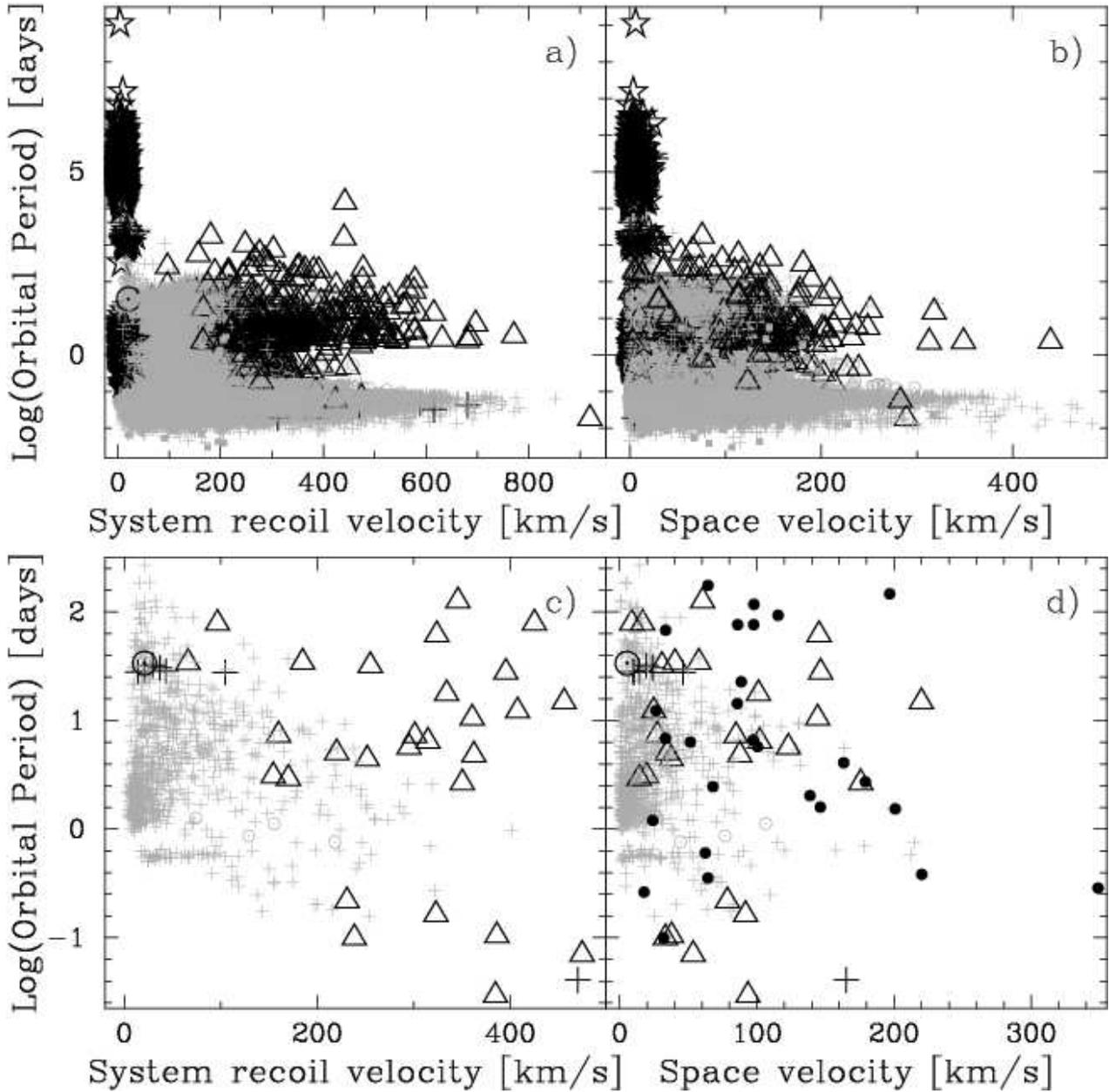}
  \caption{Parameter space of velocities (recoil and space) and
    orbital period for MSPs in Model C.
    The upper panels show all systems while the lower panels
    are restricted to include only pulsars with $B \geq 6\times 10^7~$G.
    The recoil velocities are with respect to the pre-NS binary
    CofM.
    The space velocities are with respect to the local 
    standard of rest velocity (of $\sim 220~$km s$^{-1}$: 
    Dehnen \& Binney 1998). 
    Identified are MSPs with MS star ($\odot$ symbols), giant star (solid squares), 
    WD (plusses), NS (triangles) and BH (stars) companions. 
    The large darker points represent those systems with
    eccentricities greater than $0.1$.
    The filled circles within panel d) represent 
    the $28$ MSPs with observed transverse velocities and 
    orbital periods taken from the ATNF Pulsar Catalogue.
    \label{f:fig15}}
\end{figure*}

We now examine the MSP population in velocity and 
orbital period parameter space. 
In terms of velocity we consider both the recoil velocity 
and the space velocity of the binary systems. 
The recoil velocity is defined as the change in velocity of the 
binary centre-of-mass owing to the SN explosion that 
created the NS (that went on to become the MSP). 
The space velocity is the velocity of the binary within the Galaxy 
at the time when the Galactic age is $10\,$Gyr. 
In calculating the final space velocities the solar 
motion around the Galactic centre is accounted for by removing the  
local standard of rest (LSR) velocity of 
$\sim 220~$km s$^{-1}$ (Dehnen \& Binney 1998). 
The orbital period is taken as the final orbital period at a 
Galactic age of $10\,$Gyr. 
In a similar vein to Section~\ref{s:MSP1} we examine the parameter 
space when considering all pulsars and then examine it again after 
limiting the sample population to MSPs that have 
$B \geq 6\times10^7~$G. 
The results are shown in Figure~\ref{f:fig15}. 

The recoil velocities for all MSP binaries are shown in 
Figure~\ref{f:fig15}a). 
The first item to note is the MSP-BH systems which all 
have low recoil velocities but cover a large range of final 
orbital periods.  
Such a distribution is not surprising given our detailed 
analysis of such systems in Section~\ref{s:MSP2}.
Also not surprising is the rather large recoil 
velocity range of double-NS systems. 
The typical total recoil velocity incident on such 
systems is greater than $200~$km s$^{-1}$.
We can also see from Figure~\ref{f:fig15}a) that these systems 
are likely to be eccentric rather than circular. 
For MSP-NS systems that receive large recoil velocities 
($> 450~$km s$^{-1}$) there appears to be a lower limit to the
possible final orbital period. 
The initial orbital period of these systems is very 
important in determining the evolution outcomes and the 
appearance of the final parameter space (Tauris \& Bailes 1996). 
Also, in a related manner and as discussed for MSP-BH systems in 
Section~\ref{s:MSP2}, the onset of mass transfer and the 
details of the common-envelope phase are crucial factors. 
What we find is that a significant proportion of the double-NS 
population end up with extremely small periods (and a range of eccentricities) 
and coalesce rapidly (within a few Myr after double-NS formation)
similar to that found by Belczynski, Bulik \& Rudak (2002). 
This leads to the orbital period gap observed in Figure~\ref{f:fig15}a). 
We leave further discussion on these systems for 
future work (Kiel, Hurley \& Bailes, submitted). 
Turning to the MSP-WD systems we see that these typically receive 
rather low recoil velocities with the average value being less than 
$100~$km s$^{-1}$ (much less than 
$\sqrt{2} V_\sigma$) and in accordance to previous population
synthesis results of Ramachandran \& Bhattacharya (1997),
Phinney \& Kulkarni (1994), Lyne et al. (1998) and Sun \& Han (2004).
We also see that a similar but opposite trend occurs for MSP-WDs 
as did for the MSP-NSs in that for large recoil velocity values 
there appears to be an upper limit to the possible final orbital period.  
Again this is related to the orbital evolution and in particular whether 
a system enters common-envelope evolution (and survives without coalescence) 
or not. 

Figure~\ref{f:fig15}c) shows the recoil velocity and final orbital period 
parameter space for the magnetic-field limited MSP population. 
We see that the population has been significantly thinned out. 
In particular the entire MSP-BH population has been removed as have 
the low-period MSP-WD systems. 
It is also possible to compare our findings to the results of 
Tauris \& Bailes (1996: see their Figure 2c) who followed the 
formation of MSPs using stellar and binary evolution algorithms that 
were quite advanced for their time. 
Compared to Tauris \& Bailes (1996) we find less systems with orbital 
periods greater than a day. 
However, for the MSP-WD population we observe a similar trend of 
orbital period to recoil velocity: the smaller the orbital period
the greater the range in possible recoil velocity of the system. 

In Figure~\ref{f:fig15}b) we look at the final space velocities 
and orbital periods for all MSP binaries.  
There is much similarity between the velocities 
given to each system (their recoil velocities) 
and their LSR Galactic motion.
The form of this parameter space is therefore 
governed by the same evolutionary phases that 
dictated the appearance of Figure~\ref{f:fig15}a). 
In Figure~\ref{f:fig15}d) we show the space velocity-orbital period
parameter space distribution of MSPs with 
$B \geq 6\times10^7~$G. 
Included for comparison are
pulsar proper motion observations which, 
convolved with distance estimates, give rise to 
observed transverse velocities.
From the ATNF pulsar catalogue 
(Manchester, Hobbs, Teoh \& Hobbs 2005) 
there are at present $28$ Galactic disk pulsars with spin periods
less than $0.02~$s that have measured orbital periods and 
estimated transverse velocities.
Although we do not directly compare total 
model space velocities to observed pulsar proper motions, 
useful information can still be gleaned from simple 
comparisons between the two noting
that the transverse velocities are a lower limit to 
the true space motion (although measurement errors not 
included within Figure~\ref{f:fig15}, especially in 
distance calculations, cloud this picture slightly).
Firstly, once accounting for the LSR, 
we can see that many of the model MSP systems 
travel with speeds within the typical stellar 
velocity range of approximately $\pm 16~$km s$^{-1}$ 
(as given by Dehnen \& Binney 1998). 
However, there appears to be an overabundance of model MSP 
binaries at low velocities. 
The model also fails to produce enough of the 
fast moving MSPs with large orbital periods. 
One reason for this may be the $V_\sigma = 190~$km s$^{-1}$
assumed in Model C which is lower than the value suggested 
from observations ($V_\sigma = 265~$km s$^{-1}$: Hobbs et al. 2005). 
On the other hand, Figure~\ref{f:fig15}d) shows a large range of 
space velocities for MSP-NS systems. 
Perhaps surprisingly some of these systems even have 
velocities close to the LSR, although not so surprising
according to Dewi, Podsiadlowski \& Pols (2005) who
suggest that DNSs receive small kicks. 
This is of particular interest for understanding the
nature of the double pulsar J0737-3039 (Burgay et al. 2003)
which is observed to have a transverse velocity of $30~$km s$^{-1}$
or less (Kramer et al. 2006).
Considering that the system will have experienced two supernova
events this has been taken as evidence for little or no velocity
kicks within this system.
However, Kalogera et al. (2007) have described models which
show that kick velocities of $100~$km s$^{-1}$ or more
are still possible.
Our results agree with this in that it is not necessary to make
any unusual assumptions regarding kicks in binary systems to
explain the observed velocities of systems such as J0737-3039
(Deller, Bailes \& Tingay 2009).

\subsubsection{Effects of the assumed initial scale height}
\label{s:assinitsh}

Up until now we have set a maximum of $|z_{\rm maxi}| = 75~$pc
to the initial birth height distribution of binaries, 
effectively modelling a thin disk.
We now examine the effect this has on the 
scale height calculations by extending it to 
$|z_{\rm maxi}| = 150~$pc in Model C$^{'}$.
The results are compared to Model C in Table~\ref{t:table4}. 
We find that there is no significant change in the calculated 
scale heights. 
This agrees with previous works, such as Paczynski (1990) or 
Sun \& Han (2004), who have suggested that for 
such kinematically active systems as pulsars
the initial height above the plane does not 
greatly affect the outcome. 

\subsubsection{Effects of the assumed Galactic age}
\label{s:galage}

The age of the Galaxy is an important assumption, especially
when populations of systems with large differences in 
life times are modelled together.
To address this we have Model C$^{''}$ which assumes the Galactic
age is $15~$Gyr rather than $10~$Gyr for Model C.
It is the isolated MSPs and MSP-NSs whose scale heights 
change the most appreciably in this new model compared to Model~C. 
This is primarily owing to low number statistics 
(see Figure~\ref{f:fig11} for the relative numbers of the populations). 
Otherwise it does not appear that the increase in Galaxy age 
has a significant effect on the kinematics of the MSPs. 

\subsubsection{Sufficient model resolution?}
\label{s:resolution}
Finally, to test whether our previous models have a fine enough
resolution in the initial parameter space to faithfully represent
the entire Galactic pulsar population we have extended
Model C to include $10^9$ binary systems (a factor of $100$
increase).
The only systems for which the scale height changed noticeably 
was the MSP-NS systems, which are relatively rare 
and kinematically energetic systems.
In all other respects it appears that the results for
$10^7$ systems scale reliably to larger populations.
We note here that modelling $10^9$ binary systems is equivalent
to modelling $\sim 10\%$ the mass of the Galaxy, assuming binary 
systems are of interest.
To run a model of this size takes roughly $4500$
CPU hours and even when farming the model out to $100$ processors 
(on the Swinburne 
supercomputer\footnote{http://astronomy.swin.edu.au/supercomputing/}) 
it takes almost $2~$days to complete.
Thus it is obviously an advantage when examining a variety 
of evolutionary assumptions one model at a time to be able 
to represent the Galaxy faithfully with fewer systems. 

\section{Discussion} 
\label{s:disc}

Using our newly developed  \textsc{binkin} module for integrating 
the positions of stars and binaries within the Galaxy we have worked 
through a series of models in order to understand how various options 
available in the module affect the outcomes. 
This has allowed us to develop a favoured model -- our Model C. 
In doing this we have used pulsar populations as our yardstick, 
computing scale heights, radial and velocity distributions, and 
orbital characteristics (in the case of binary systems) with limited 
comparison to observations. 
What we have done is to make predictions in all these areas about 
the particulars of the Galactic pulsar population, assuming that 
all pulsar systems can be observed. 
Of course this is not the case in reality and our model results cannot 
truly be confronted by observations until we include selection effects 
in our modelling. 
This will be completed when we add our next and final module \textsc{binsfx} 
to our population synthesis code. 
As such we leave a discussion of the necessary selection effects that need 
to be considered and their treatment to an upcoming paper focussed on the 
\textsc{binsfx} module (Kiel, Bailes \& Hurley, in preparation). 
This paper will include features such as the predicted 
pulsar $P$-$\dot{P}$ diagram for distinct regions in the Galaxy 
(following on from our investigation of this diagram in Paper~I 
in terms of binary evolution parameters). 
Below we discuss future additions to the \textsc{binpop} and 
\textsc{binkin} modules in relation to pulsar evolution as well 
as caveats to our current findings. 

\subsection{Accretion induced collapse formation of neutron stars}
\label{s:AICdisc}

Further analysis of our MSP populations shows that some 
of the NSs which go on to become MSPs in our models 
are formed from the accretion induced collapse (AIC) of WDs 
(Canal \& Schatzman 1976; Nomoto \& Kondo 1991). 
In the scenario of Nomoto \& Kondo (1991) an O-Ne-Mg WD 
accretes enough material to reach the Chandrasekhar mass, 
the maximum mass possible for a WD to support itself, 
and collapses to form a NS. 
To date we have allowed NSs formed in this way to receive 
velocity kicks in the same manner as for NSs formed in 
core-collapse SNe. 
Generally, if the binary system remains bound an AIC NS 
will continue to accrete material after the SN, a SN which 
induced an eccentricity into the orbit.
This formation pathway produces a substantial number of 
MSP-WD and MSP-MS systems with eccentricities greater than
$0.1$, that reside within the Galaxy.
These systems highlight the importance of a correct mass-transfer 
treatment for eccentric binaries (see Paper~I and 
Bonacic-Marinovic, Glebbeek \& Pols 2008).
This is something which is not currently accounted for in our models 
(we make use of equations which assume the orbit is circular: 
see HTP02) and it most likely will affect the production and 
visibility of these systems.
Of course, if the AIC systems were not given any velocity kick,
(as has been modelled previously: HTP02), or a much lighter kick 
(as latest models suggest: Dessart et al. 2006), then not only would
there be many more AIC MSP systems 
but they would all have a greater possibility of residing in our 
Galactic target area and the population scale height would be lowered. 
They would also typically have smaller eccentricities.

Although we do not deal directly with low-mass X-ray binaries 
within this work it is possible for us to 
compare the observed scale heights of such systems -- 
which suffer from less selection effects than pulsar 
observations -- with our model MSP-MS scale height 
calculations.
This is assuming that low-mass X-ray binaries are the 
progenitors of MSP-MS systems.
Grimm, Gilfanov \& Sunyaev (2002) found that 
Galactic field low-mass X-ray binaries have a scale 
height of $\sim 0.410~$kpc. 
Interestingly enough, as shown in Table~\ref{t:table4}, 
our models over estimate this by almost a factor of two. 
Such an outcome may be another implication that 
AIC NSs receive less momentum at birth than standard NSs 
formed from core-collapse SNe. 
However, it is not clear that MSPs that result from AIC 
NSs can be linked to an observable low-mass X-ray binaries 
phase (Hurley, Ferrario, Wickramasinghe, Tout \& Kiel, in prep).

\subsection{Electron capture supernovae}

Another evolutionary scenario related to NS formation and velocity 
kicks is core-collapse electron capture SNe.
This was discussed and modelled in Paper~I and has also been 
accounted for in other population synthesis works 
(e.g. Ivanova et al. 2008).
Briefly, core collapse electron capture SNe are thought to arise
when electrons are captured onto Mg atoms, depleting the
electron force in an O-Ne-Mg stellar core of sufficient mass 
($1.4-2.5~M_\odot$: Nomoto 1984) which is produced by initial 
progenitor masses in the range of $8 - 12~M_\odot$ 
(although this mass range is model dependent: Podsiadlowski et al. 2004).
The likelihood that a star born within the $8-12~M_\odot$ limit will
evolve to have an O-Ne-Mg core mass between
$1.4-2.5~M_\odot$ increases if the progenitor is able to interact with
a companion and lose its outer hydrogen envelope,
rather than evolve in an isolated environment 
(Podsaidlowski et al. 2004). 
Therefore, binary population synthesis is ideal for examining the likelihood 
and outcomes of such events. 
The resultant electron capture SN energy yield is low, 
sufficient to cause the explosion but not enough to 
impart any large velocity to the proto-NS 
(Kitaura, Janka \& Hillebrandt 2006).
Paper~I found that the final MSP spin period and
spin period derivative parameter space was altered 
when electron capture SNe were included.
Less pulsar binary systems were disrupted, owing to 
the small momentum imparted during the SN, causing 
more MSPs to be produced. 
It is also reasonable to expect that including electron capture SNe 
in the \textsc{binkin} models, with SN kicks drawn from a distinct 
distribution with a smaller velocity dispersion than for standard NSs, 
will lead to a reduction in the pulsar scale heights. 
This is a feature that will be fully explored in future models so that 
the impact of the electron capture SNe process on binary evolution 
outcomes and the resultant Galactic kinematics of pulsar populations 
can be quantified. 

\subsection{MSP-BH systems}

In our models we have assumed no SN velocity 
kick is given to BHs and have found that MSP-BH systems 
reside close to the Galactic plane.
If BHs were instead to receive a SN kick selected from 
the same distribution as NSs then it is clear that the scale heights 
of populations including BHs would increase (Voss \& Tauris 2003). 
However, it is not so obvious that the scale heights would be 
similar to that of the equivalent NS populations (Pfahl et al. 2005). 
In particular, MSP-BH systems (and their progenitors) 
will be heavier on average than MSP-NS systems 
(and their progenitors) and the 
more massive systems will require a greater momentum to
reach the same velocities as less massive systems. 
As such MSP-BHs for example, could still have a significant 
difference in their resultant scale height to that of MSP-NSs 
even when both populations receive kicks from the same 
distribution. 
We would also expect the number of BH binary systems to 
decrease. 
Most likely it would be the MSP-BH systems with a large 
orbital periods prior to BH formation (systems with initial 
orbital periods greater than $\sim 100~$days) which 
would be depleted.
It is these systems that are not produced in the models of
Pfahl et al. (2005) who assume SN kicks occur on nascent BHs.
However, we must bear in mind that the final BH masses are 
calculated assuming that material ejected in the SN falls back 
on to the BH. 
There is less mass-loss associated with BH formation than 
for NSs and this means supernova induced binary disruption 
is less likely during BH formation (in the case of equivalent 
kick velocities).  

We note that when discussing the MSP-BH population 
(or any of our model MSPs) we are defining a rapidly rotating NS 
to be an MSP based solely on its spin period. 
If instead we also include consideration of the magnetic field 
strength of these NSs then the nomenclature may be misleading, 
especially if we are interested in observable MSPs. 
It turns out that all of the NSs in our model MSP-BH systems 
have magnetic fields residing on, or very close to, the assumed bottom 
magnetic field limit of $6\times10^7~$G 
(Paper I; Zhang \& Kojima 2006). 
Previously (Paper I, Figure~\ref{f:fig12} and 
Figure~\ref{f:fig15}), we have assumed that any 
NS with a magnetic field less than this limit 
cannot accelerate the electrons 
in its atmosphere to produce the observed pulsar 
beam and as such is not observable as a pulsar. 
Therefore, if our assumptions regarding accretion 
on to NSs and how this translates to magnetic field 
decay are correct then we have a lot of trouble 
producing observable MSP-BHs. 
Future observations of such systems will help greatly in 
constraining our evolutionary assumptions.

\subsection{Initial distributions}
\label{s:dis2}

In our models we have assumed 
a maximum birth height off the plane, $|z_{\rm maxi}|$, 
of either 75 or $150\,$pc.  
Consistent with Paczynski (1990) and Sun \& Han (2004)
no significant variations of the MSP population scale heights 
were found when varying this parameter. 
This suggests that the results are robust to changes in $|z_{\rm maxi}|$ 
as long as a sensible choice is made. 
The majority of OB star formation has been shown by 
de Wit, Testi, Palla \& Zinnecker (2005) 
to occur within $|z| \sim 200~$pc of the Galactic plane so 
choices within this range, such as for our models, would 
seem reasonable. 
In the future it will be interesting to probe the effects
of assuming a radial dependence in $|z_{\rm maxi}|$
on the final pulsar population distributions.
This may even be tied in with examining the effect 
of assuming bursts of star formation throughout the 
age of the Galaxy and accounting for Galactic arms 
when initiating the birth positions.
This final point has previously been suggested as an 
important feature to incorporate into population
synthesis models (Faucher-Giguere \& Kaspi, 2007).

We found that the Yusifov \& Kucuk (2004) initial radial 
pulsar birth distribution gave the best 
fit of our models to observations.
This distribution was based on observations of HII regions 
within the Galaxy.
However, it failed to reproduce the peak of the 
observable radial distribution -- which the Paczynski (1990)
initial radial distribution succeeded in reproducing.
Yusifov \& Kucuk (2004) recognised that their relation is
approximate and suggested that a detailed analysis 
between models and observations of 
pulsar velocities and Population I stellar positions
was required to develop a more realistic distribution.
We are in a position to do this and as a result can suggest 
that the initial pulsar birth distribution 
of Yusifov \& Kucuk (2004) perhaps be shifted towards smaller Galactic radii to
peak at the inner HII peak ($\sim 4.0-4.5~$kpc) depicted 
in Figure 3 of Paladini, Davis \& DeZotti (2004).

\subsection{Galactic model potentials} 

Even though our favoured model (Model C) utilises the Pac90 
form of the Galactic gravitational potential we are in no way able 
to distinguish between this and the KG89 model as a more 
suitable representation of the Galactic potential. 
Both give similar pulsar population scale height results which 
is not surprising given their similarities as shown in Figure~\ref{f:fig4}. 
The form of the Xue08 potential is clearly distinct from the other 
two models, especially within the inner $1\,$kpc of the Galaxy 
(where Xue08 employ an extrapolation of their measurements), 
and leads to markedly increased scale heights. 
On this basis we do not favour use of the Xue08 potential. 
However, we are not currently in a position to make strong 
conclusions in this area, especially when many previous
pulsar, NS and X-ray binary population synthesis works
(such as Paczynski 1990, Lorimer et al. 1993, 
Belczynski, Bulik \& Rudak 2002, Sun \& Han 2004 and 
Zuo, Li \& Liu 2008) have used different Galactic 
gravitational potentials and their results compare 
well to observations.
We note that Sun \& Han (2004) comment that it is unclear whether the 
Milky Way has a peak in the gravitational potential at
small Galactic radii (as present in the Pac90 and KG89 models). 

A possibility in the future is to extend the Galactic gravitational potential 
analysis to consider Modified Newtonian Dynamics (MoND). 
Such an approach has already been taken by Wu et al. (2008),  
who compare MoND with cold dark matter models,  
and Zuo, Li \& Liu (2008) who make use of MoND potentials to 
conduct population synthesis of X-ray binaries.

\subsection{Close double compact systems and gamma ray bursts}

In this work we have focussed on pulsars and looked in detail at MSP systems. 
However, the models can also be extended to explore the formation of 
close double compact systems (NS-NS, BH-BH and NS-BH systems) 
in detail. 
The kinematics of these systems is of interest because of their link 
to gamma-ray bursts and, in particular, recent observations of the 
distances at which gamma-ray bursts appear to occur from their 
(assumed) host galaxy (Bloom, Kulkarni \& Djorgovski 2002). 
Our combined  \textsc{binpop} and \textsc{binkin} modules can provide 
model estimates for the projected distances from their host galaxy at 
which double compact systems coalesce and document the kinematic 
evolution of these systems in general. 
This will be the focus of a companion paper 
(Kiel, Hurley \& Bailes, submitted). 

\section{Summary}
\label{s:kinconc}

We have examined in depth the Galactic dynamics and 
population characteristics (owing to stellar, binary and kinematic 
evolution) of pulsars. 
%
Our main findings, reconfirming and updating many areas of 
pulsar evolutionary physics, can be summarised as follows 
(noting that overlap with previous work is detailed in 
Section~\ref{s:results}): 
\begin{itemize} 

\item When using a peaked radial distribution for the 
birth locations of binaries, the population of pulsars that 
arises from these binaries also follows a peaked distribution 
where the location of the peak moves inwards in radius by 
as much as $0.5\,$kpc as the population evolves. Also, 
compared to the birth distribution, the initial shape is 
preserved inward of the peak but the distribution becomes 
more extended in the outer regions. 

\item Starting with a uniform initial distribution of binaries 
cannot produce a final pulsar distribution that is peaked 
away from the centre of the Galaxy and therefore does not 
compare well to observations of pulsar locations which indicate 
a deficit of pulsars towards the Galactic centre. 

\item The form of the Galactic potential does not produce 
significant differences in the final radial distribution of pulsars 
but can lead to noticeable differences in the calculated scale heights of 
pulsars. 

\item As the pulsar population ages the peak of its velocity 
distribution moves to lower velocities. 
The velocity dispersion of this distribution (assuming a Maxwellian) 
almost halves over a period of $10\,$Gyr. 
The shape of the velocity distribution is significantly 
affected by the the inclusion of binary evolution 
-- this produces a more sharply peaked distribution. 

\item Similar to observations we find that the majority 
of standard pulsars are isolated and that these dominate 
the statistics of the pulsar scale height calculations. 

\item Isolated pulsars have a 
greater scale height than binary pulsars except in cases 
where large velocity kicks are applied to the population 
resulting in many isolated pulsars being lost from the Galaxy 
and hence from the scale height calculations. 

\item Isolated MSPs have greater scale heights than binary MSPs 
(by as much as a factor of two) however, 
limiting the region of the Galaxy considered (in terms of height 
off the plane) does reduce the difference in these scale heights 
and brings them more in line with what observations suggest. 

\item We find that 99\% of MSPs are in binary systems when we 
only consider SN disruption as a pathway for creating isolated MSPs. 
This does not agree with the observed MSP population. 
If we include a simple ablation model we find instead that 64\% 
of MSPs are in binaries which adequately matches the observed mix. 
Furthermore, accounting for ablation gives similar scale heights for 
isolated and binary MSPs. 

\item MSP systems with NS companions can receive large recoil velocities. 
There is a large scatter in the resulting peculiar motions of MSP-NS binaries 
and it is possible for such systems to found with low peculiar motion.

\item The scale heights of the MSP-MS and MSP-WD binary populations 
are very similar and follow similar radial distributions. 
These scale heights are larger than that of the observed 
low-mass X-ray binary population in the Galaxy (often thought to 
be the precursors of MSP-MS binaries). 
However, many of the model MSPs in binary systems are formed from the 
accretion-induced collapse of a WD which, if given smaller kicks than for 
standard NSs at birth, would reduce the model scale heights. 

\item MSPs with WD companions are the most common of the binary MSPs. 
This is followed by MSP-MS, MSP-BH and MSP-NS binaries, respectively. 

\item Restricting the model MSP population to only include MSPs 
with magnetic fields greater than $6 \times10^7\,$Gauss 
drastically reduces the number of systems and changes the way that 
the population is distributed. 
This suggests that the underlying pulsar distribution of the Galaxy may differ
greatly from the observed sample.

\end{itemize} 

One future goal of pulsar astronomy is the detection
of a pulsar orbiting a black hole, and in terms 
of placing constraints upon general relativity a 
millisecond pulsar in a close orbit around a black hole
would be an especially exciting observation.
We find three distinct evolutionary pathways
which result in the formation of MSP-BH systems. 
These pathways produce three distinct MSP-BH populations 
in terms of orbital period: those with periods of $10\,$day or less, 
those with periods of about $1\,000\,$days, and those with periods 
of $10\,000\,$day or greater. 
The short and long period populations are the most numerous 
and only the short-period systems are found further than $1\,$kpc from 
the Galactic plane. 
We find that owing to the amount of material accreted by the MSPs 
in our model MSP-BH binaries that the magnetic field decays 
below $6 \times10^7\,$Gauss. 
This possibly suggests that we are overestimating the rate of 
accretion-induced magnetic field decay in our evolution model 
-- the observation of a MSP-BH binary would confirm this possibility. 

We emphasise to the reader that we are not presenting any of the 
models in this paper as a definitive representation of the true 
Galactic pulsar population. 
The uncertainty involved in the many parameters contained 
within \textsc{binpop} and \textsc{binkin} does not allow this. 
Moreover, because we do not consider selection effects in our 
model Galaxy we cannot at this stage make definitive comparisons 
to observations as the possibility exists that the observed population  
may be biased in some manner. 
Lommen et al. (2007) suggest that observations of MSPs 
may preferentially detect binary MSPs because the isolated MSPs 
may be less luminous than their binary cousins.
The intrinsic luminosity of pulsars is not something examined
in this body of work. 
However, it will be discussed in detail in future work where 
selection effects are calculated within our upcoming \textsc{binsfx} module 
(Kiel, Bailes \& Hurley, in prep).
Supplementing our current pulsar population synthesis with selection effects 
will allow additional evaluation of the evolutionary codes
and their scientific outcomes.
It will also allow us to guide further surveys by selecting
regions of the sky best suited for the specific pulsar survey
and/or telescope of interest.
Therefore \textsc{binsfx} will provide a powerful tool with which to 
constrain the theory and modelling of stellar, binary and Galactic 
kinematic evolution.
Further constraints could be placed on binary evolution
if population synthesis studies are extended to include 
additianal stellar populations \textit{and} their appropriate 
selection effects.
For now, however, we are well on our way to producing a 
comprehensive treatment of pulsar population physics.

\section*{Acknowledgments}
PDK and JRH wish to thank Matthew Bailes and the referee 
for helpful comments and suggestions.
PDK also thanks Swinburne University of Technology for a
PhD scholarship.

\appendix
\section{}
\label{s:isomsp}
The time taken for a star to be ablated by a MSP can 
be approximated by taking the irradiated luminosity
onto the companion, which is of order
\begin{equation}
L \sim \frac{\Delta E}{\Delta t} \sim 4.4\times 10^{32} \rm{erg s^{-1}},
\end{equation}
(Tavani, 1992) and equating $\Delta E$ with the change in binding 
energy of the companion, 
\begin{equation}
E_{bind} = \frac{GM_\star^2}{R_\star},
\end{equation}
Solving $\Delta t$ for the threshold mass of a 
$0.02~M_\odot$ star (Tavani 1992) gives 
$\Delta t = 5.5~$Myr, while for a $0.01~M_\odot$ 
star $\Delta t = 2.7~$Myr.

\bsp

\label{lastpage}

\end{document}